\documentclass{article}
\usepackage{arxiv}

\usepackage{graphicx}
\usepackage{mathptmx}      % use Times fonts if available on your TeX system
\usepackage{latexsym}
\usepackage[misc,geometry]{ifsym}  % letter symbol to indicate corresponding author
\usepackage{hyperref}				% url formatting
\usepackage{amsmath}				% helpful math stuff
\usepackage{amssymb}
\usepackage{tabularx}
\usepackage{booktabs}				% better lines for tables
\usepackage{threeparttable}			% footnotes in table
\usepackage[utf8]{inputenc}
\usepackage[T1]{fontenc} 			% \k{e} special character
\usepackage[numbers]{natbib} 	% citation management
\usepackage{multirow} 				% span rows in a table
\usepackage{array}					% for using c & p together
\usepackage{xcolor}                 % for coloring text
\usepackage[nolist,nohyperlinks]{acronym}
\acrodef{ET-HMD}[ET-HMD]{tethered system based on the HTC Vive}
\acrodef{VR}[VR]{virtual reality}
\acrodef{HMD}[HMD]{head-mounted display}
\acrodef{VAC}[VAC]{vergence-accommodation conflict}
\acrodef{RMS}[RMS]{sample-to-sample root mean square error}
\acrodef{SD}[SD]{standard deviation}
\acrodef{BCEA}[BCEA]{bivariate contour ellipse area}
\acrodef{MAD}[MAD]{median absolute deviation}
\acrodef{EOG}[EOG]{electrooculography}
\acrodef{VOG}[VOG]{video-oculography}
\acrodef{DPI}[DPI]{dual-Purkinje-image}
\acrodef{MSC}[MSC]{manufacturer-supplied calibration routine}
\acrodef{USC}[USC]{user-supplied calibration routine}
\acrodef{ISI}[ISI]{intersample interval}
\acrodef{R2}[$R^2_{adj}$]{adjusted R-squared}

\makeatletter
\renewcommand*\AC@acs[1]{%
    \expandafter\AC@get\csname fn@#1\endcsname\@firstoftwo{#1}}
\makeatother

\newcommand{\degree}{^{\circ}}

\newcommand{\mnsd}[2]{#1\,$\pm$\,#2}

\newcolumntype{P}[1]{>{\centering\arraybackslash}p{#1}}

\makeatletter
\newcommand{\ssymbol}[1]{$^{\@fnsymbol{#1}}$}
\makeatother

\DeclareRobustCommand{\VON}[3]{#2} % set up for citation at the start

\makeatletter
\def\input@path{{Sections/}{Tables/}}
\makeatother

\author{
	Dillon J.~Lohr \\
	Department of Computer Science \\
	Texas State University \\
	San Marcos, TX 78666, USA \\
	\texttt{djl70@txstate.edu} \\
	\And
	Lee Friedman \\
	Department of Computer Science \\
	Texas State University \\
	San Marcos, TX 78666, USA \\
	\texttt{l\_f96@txstate.edu} \\
	\And
	Oleg V.~Komogortsev \\
	Department of Computer Science \\
	Texas State University \\
	San Marcos, TX 78666, USA \\
	\texttt{ok11@txstate.edu}
}
\title{%
    Evaluating the Data Quality of Eye Tracking Signals from a Virtual Reality System: Case Study using SMI's Eye-Tracking HTC Vive %
}
\begin{document}

\maketitle

\begin{abstract}
    We evaluated the data quality of SMI's tethered eye-tracking head-mounted display based on the HTC Vive (\acs{ET-HMD}) during a random saccade task. %
    We measured spatial accuracy, spatial precision, temporal precision, linearity, and crosstalk. %
    We proposed the use of a non-parametric spatial precision measure based on the \acf{MAD}. %
    Our linearity analysis considered both the slope and \acl{R2} of a best-fitting line. %
    We were the first to test for a quadratic component to crosstalk. %
    We prepended a calibration task to the random saccade task and evaluated 2 methods to employ this user-supplied calibration. %
    For this, we used a unique binning approach to choose samples to be included in the recalibration analyses. %
    We compared our quality measures between the \acs{ET-HMD} and our EyeLink~1000 (SR-Research, Ottawa, Ontario, CA). %
    We found that the \acs{ET-HMD} had significantly better spatial accuracy and linearity fit than our EyeLink, but both devices had similar spatial precision and linearity slope. %
    We also found that, while the EyeLink had no significant crosstalk, the \acs{ET-HMD} generally exhibited quadratic crosstalk. %
    Fourier analysis revealed that the binocular signal was a low-pass filtered version of the monocular signal. %
    Such filtering resulted in the binocular signal being useless for the study of high-frequency components such as saccade dynamics. %
\end{abstract}


\acresetall

\section{Introduction}
We recently acquired an eye-tracking \ac{VR} \ac{HMD}: SMI's \ac{ET-HMD}.
Our overall goal of this work was to fully and carefully characterize the data quality produced by this device.
%Many similar devices are already, or will soon be, on the market (e.g., FOVE,\footnote{\url{https://www.getfove.com/}} VIVE Pro Eye,\footnote{\url{https://vr.tobii.com/products/htc-vive-pro-eye/}} Varjo VR-1\footnote{\url{https://varjo.com/vr-1/}}).
Many similar devices are already, or will soon be, on the market (e.g., FOVE, VIVE Pro Eye, Varjo VR-1).
Our methods and results may be of interest to other researchers working with such devices.

\begin{figure}[htp]
\centering
\includegraphics[width=\linewidth,trim={0, 0, 0, 0.8cm},clip]{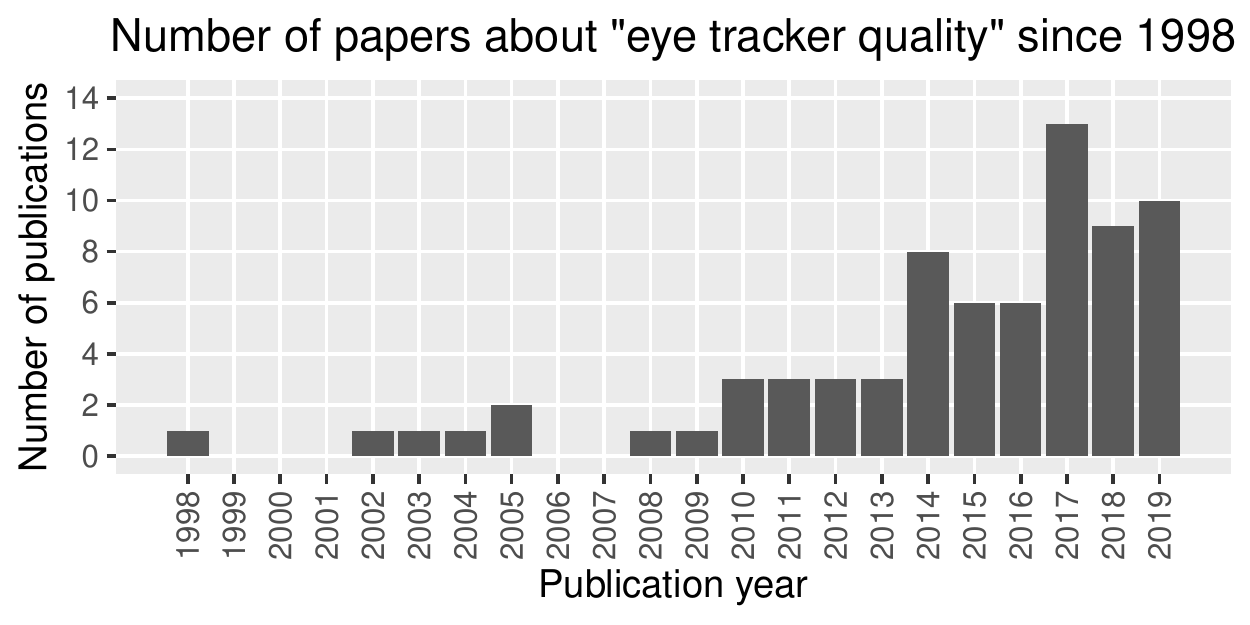}
\caption{
	The number of articles in the PubMed database containing the words ``eye,'' ``tracker,'' and ``quality'' as of November 7, 2019.
	Papers are grouped by the year they were added to PubMed.
	Partial data for 2019 (approximately 10 months).
	% Articles from 2019 are shown in a lighter gray, above 2018.
}
\label{fig:pubmed}
\end{figure}

One important reason for a detailed analysis of the data quality of eye trackers is that the manufacturers' published specifications are often not achievable in practice~\citep{Blignaut2014,Holmqvist2011,Nystrom2013}.
Studies of eye movements in reading need to know, as accurately as possible, which exact word is being read and even which letter is being fixated~\citep{Rayner1998,Reichle2003}.
The use of eye position as a selection device for options on a web page also seems like an example where very accurate tracking would be required.
In the context of oculomotor biometrics (the effort to identify individual humans based on their eye movements)~\citep{Friedman2017,Holland2011}, improved performance of the eye-tracking device is likely to lead to an improved estimate of inter-human variation.
The need for increased data quality may be the motivation behind the recent increase in the number of publications including the words ``eye,'' ``tracker,'' and ``quality'' (Fig.~\ref{fig:pubmed}).

When we refer to data quality, we are specifically referring to measures of spatial accuracy, spatial precision, temporal precision, linearity, and crosstalk, each defined in Table~\ref{tbl:definitions} below.
\begin{table*}[htp]
	%\centering
	%\captionsetup{font=normalsize}
	\caption{Definitions of Data Quality Terms Used in This Paper}
	\label{tbl:definitions}
	\begin{tabularx}{\textwidth}{lX}
		\toprule
		Term & Definition \\
		\midrule
		% Spatial accuracy & The difference between the true gaze position and the measured gaze position \\
		Spatial accuracy & The ability of the eye tracker to correctly measure the gaze position \\
		Spatial precision & The ability of the eye tracker to reliably reproduce a gaze position measurement \\
		Temporal precision & The ability of the eye tracker to produce new gaze position measurements at a consistent rate \\
		Linearity & The consistency of spatial accuracy across the spatial range of the recording device \\
		Crosstalk & The extent to which movement in one direction (horizontal or vertical) artifactually influences movement in the orthogonal direction \\
		\bottomrule
	\end{tabularx}
\end{table*}
\citet{Hooge2018} argued in favor of using the term ``systematic error'' instead of ``spatial accuracy'' and the term ``variable error'' instead of ``spatial precision.''
Their view was that it is counterintuitive for a \textit{high} accuracy to be characterized by a \textit{small} systematic error, and likewise for a \textit{low} precision to be characterized by a \textit{large} variable error.
Indeed, it may be confusing, especially for new researchers, that a \textit{low} spatial accuracy actually means that the calculated value is \textit{large}.
Although we agree with the semantic argument by \citet{Hooge2018}, we will use the terms ``spatial accuracy'' (or just ``accuracy'') and ``spatial precision'' (or just ``precision'') to be consistent with the majority of the literature.

Regarding spatial precision, \citet{Hornof2002} suggested that:
\begin{quote}
    ``Variable error can be seen in the spread (or \textit{dispersion}) of the recorded gazepoints around the actual fixation maintained by the human.
    Systematic error is the disparity (or \textit{drift}) between the average gazepoint location and the actual fixation maintained by the human.
    Variable error indicates a lack of \textit{precision}.
    Systematic error indicates a lack of \textit{accuracy}'' (p.~592).
\end{quote}
We disagree with a formulation of precision as a measure of spread and accuracy as a measure of distance.
It defines accuracy and precision as two fundamentally different types of things.
We prefer to think of both accuracy and precision as average-distance measures with different reference points.
% Statisticians use the term ``central tendency'' to describe the mean (or median).
Accuracy should be defined as the mean (or median) of the distances between each sample and the target position.
Precision should be defined as the mean (or median) of the distances between each sample and the central tendency of all samples.
That is, the reference point for accuracy is the target position, while the reference point for precision is the central tendency of the samples.

% TODO: Large space here, possible to fix?

\subsection{Historical efforts to assess eye-tracker quality}
Assessment of the accuracy of eye-movement measuring equipment is probably as old as the measurement of eye movements.

% Maybe split by image-based and non-image-based
In the first paper to employ a corneal reflection and the first paper to make reasonably accurate estimates of the average velocity of saccades, \citet{Dodge1901} employed a photographic method to measure eye movements. 
The method involved recording eye movements on a photographic plate that was falling at a rate which was constant within each recording but varied from recording to recording.
Using a pendulum as a time source, \citeauthor{Dodge1901} were able to determine that the average error in temporal precision was 0.5~ms.
In the paper introducing the scleral search coil method, \citet{Robinson1963} evaluated his new system and found that the device had an accuracy and linearity of about 2~percent of full scale, a spatial resolution of 15~seconds of arc, and a bandwidth of 1000~Hz.
\citet{Merchant1967}, in one of the earliest reports to use the \ac{VOG} method to estimate eye position using pupil position and corneal reflection position, found his device to have a \ac{RMS} noise level of $0.3\degree{}$ and channel crosstalk of 20\% maximum.
However, \citeauthor{Merchant1967}'s system was substantially nonlinear with respect to eye position.
In the paper introducing the \ac{DPI} tracking method, \citet{Cornsweet1973} reported that the device had a spatial resolution and spatial accuracy of about 1~minute of arc.
\citet{Collewijn1975} developed a more comfortable scleral search coil method and reported that the noise level of the system at maximum sensitivity was equivalent to about 0.5~minutes of arc.
In an early general discussion of the topic of recording quality, \citet{McConkie1981} encouraged standardization in the reporting of data quality for eye movement research.
He also described a method of measuring drift in a recording device with an artificial eye.
\citet{Reulen1988} developed an advanced infrared limbus tracker and provided an early example of a reasonably comprehensive performance evaluation, including measures of spatial resolution, spatial precision, linearity, and crosstalk.
Drift was assessed qualitatively.

\subsection{Current state of the field}
%Since these seminal works, there have been a lot of new developments.  
\subsubsection{Spatial accuracy}
There seems to be little debate regarding the definition of spatial accuracy (see Table~\ref{tbl:definitions}), although there is some variability in the exact term (e.g, ``systematic error'' or ``offset error'' \citep{Hooge2018,Hornof2002,Vadillo2015}) used to label this quality.

The first step in the measurement of accuracy is the determination of which data points to include in the calculation, and there are various approaches to this.
For example, \citet{Blignaut2014a} ignored the first 1000~ms of data after presenting a target, then used the next 500~ms.
\citet{Akkil2014} had two parts to their study: a system-controlled routine and a participant-controlled routine.
For the system-controlled routine, \citeauthor{Akkil2014} ignored the first 500~ms of data after presenting a target and used the next 500~ms of data.
For the participant-controlled routine, they identified the moment when the mouse cursor made the final approach to the target and collected roughly 500~ms of data.
\citet{Blignaut2012a} used data in the period immediately before a mouse-up event during a participant-controlled routine, with the duration of the period typically between 150-300~ms.
\citet{Kasprowski2014} ignored the first 700~ms of data after presenting a target, then used the next 1100~ms.
\citet{Blignaut2014} used 250~ms of data preceding a mouse click.
\citet{Nystrom2013} ignored the first 400~ms of data after presenting a target, then used multiple criteria, including temporal and spatial criteria, for determining which samples to include in their calculation.

There is also variation in the quantity and position of targets used.
The most commonly used target arrangement is a square grid containing 9 to 25 points \citep{Kasprowski2014}, but even these somewhat standard arrangements vary in their visual angle span.
On one end of the spectrum, \citet{Akkil2014} used a 4-point arrangement with each point located 20\% of the screen dimension away from the corners.
On the other end, \citet{Blignaut2014} employed a 40-point arrangement (8 columns and 5 rows) spanning $38.4\degree{}$ horizontally and $24\degree{}$ vertically.

\subsubsection{Spatial precision}
Aside from the obvious benefit of precise measurement, the spatial precision of a device can influence experimental results from human studies such as estimated fixation durations \citep{Holmqvist2012,vanRenswoude2018}, event detection \citep{Nystrom2013}, or investigating imperfections in the oculomotor system \citep{Nystrom2013}.
Spatial precision can be measured with an artificial eye \citep{Liston2016,Reingold2014,Tobii2012,Wang2017} or from a human recording \citep{Blignaut2014a,Liston2016,Tobii2012,Wang2017}, but \citet{Holmqvist2012} claim it is better to use both artificial eyes and human subjects when assessing spatial precision.
Using an artificial eye provides a spatial precision measurement without the complication of human oculomotor noise \citep{Holmqvist2012}, including drift, tremor, and microsaccades \citep{Leigh2006,McCamy2013}.
However, as noted by \citet{Holmqvist2012}, artificial eyes do not have the same iris, pupil, and corneal reflection features as human eyes.
Also, human eyes can vary greatly (e.g., eye color, pupil size, corneal size, degree of corneal bulge).
Therefore, testing with real eyes can provide precision data for a realistic population of potential test subjects.

Spatial precision is frequently measured either as \ac{RMS} or as an estimate of dispersion among a set of samples \citep{Holmqvist2012,Tobii2012}.
\citet{Holmqvist2012} show how \ac{RMS} measures are more resistant to vibrations in the environment than dispersion as assessed with a \ac{SD}.
However, \citet{Blignaut2012} state that a disadvantage of using \ac{RMS} is that it varies with sampling rate.
Given the time-series nature of the data stream, there is an implied temporal dependency in the calculation of \ac{RMS}.
That is, \ac{RMS} is a function of the ordering of the samples.
If one randomly shuffled the sample positions, one would in all likelihood get a different \ac{RMS}.
% They also note a disadvantage of using the \ac{SD} of a set of samples, where the precision will be close to zero if the samples are positioned more-or-less equidistant from the centroid and Euclidean distance is used as the distance measure.

In a more thorough overview of spatial precision measures, \citet{Holmqvist2011} provided eight additional measures of spatial precision.
\citet{Blignaut2012} briefly critiqued these eight measures, rejected four of them, considered three to be overly complex and difficult to interpret, and provided a detailed evaluation of \ac{RMS}, \ac{SD}, and \ac{BCEA}.
\ac{BCEA} defines the area of an ellipse that encompasses some proportion of a set of points \citep{Holmqvist2011}.
\citet{Blignaut2012} reduced \ac{BCEA} to a one-dimensional value they called r(BCEA) by approximating the ellipse with a circle.
They proposed that the use of r(BCEA) is intuitive, independent of sampling rate (unlike \ac{RMS}), and independent of the arrangement of samples within a fixation.

\subsubsection{Temporal precision}
For most eye-movement recording devices, the nominal interval between samples is measured in milliseconds, and frequently only the nominal sampling rate is available to the user.
This is certainly the case with our EyeLink~1000 (SR-Research, Ottawa, Ontario, CA), which has a nominal sampling rate of 1000~Hz and, according to staff at SR Research, produces samples precisely 1~ms apart.
%In this scenario, there is no evidence that variability exists in the sampling rate.
The \ac{ET-HMD} has a nominal sampling rate of 250~Hz, but it provides timestamps for samples with nanosecond precision.
(We are unaware of how common such precise timestamps are.)
When we use the term ``temporal precision,'' we mean the variation of the nominal sampling rate estimated by sub-millisecond timestamps.
This definition of temporal precision was used by \citet{Abdulin2019} (using timestamps precise to $10^{-7}$ seconds) in the evaluation of their custom eye-tracking device.
As we will discuss in Sect.~\ref{sec:definitions-temporal}, others use this term in a different context, befitting gaze-contingent research.
% Others use this term in a different context, befitting gaze-contingent research.\footnotemark

% \footnotetext{
%     \citet{Holmqvist2011} use the term \textit{temporal precision} to refer to the variance in system latencies, where system latency is the average delay from an actual eye movement until the recording system signals that a movement has occurred.
%     For example, this definition is used by \citet{Reingold2014}, who employs an artificial saccade generator to measure the variance in ``system end-to-end delay'' (i.e., system latency).
%     Although we see the relevance of this measure in a real-time situation, all of our analyses of eye-movement data is offline.
%     For us, this sense of temporal precision is irrelevant to our situation.}
%     % Light moves from the eye to the camera at the speed of light, and the time to process the image and compute a position is irrelevant.

\subsubsection{Linearity}
Several studies have noticed that spatial accuracy can be a function of target position \citep{Blignaut2012a,Blignaut2014,Hornof2002,Nystrom2013,Reulen1988}.
Since at least \citeyear{Young1975} \citep{Young1975}, this has often been referred to as ``linearity.''
Although this term for a dependence of accuracy on target position has not always been referred to as linearity, the relationship has been studied repeatedly in the literature.

\citet{Reulen1988} used the term linearity and specifically measured it by fitting a line to a scatter plot relating measured eye position to target position (both horizontal and vertical).
%Residuals from these lines were expressed as a percentage of the position range ($56\degree{}$ horizontal and $24\degree{}$ vertical).
%Linearity was expressed as the maximum residual (in percent).
Residuals from these lines were expressed as a percentage of the position range.
Linearity was expressed as the maximum residual.
For their device, linearity was 3\% for the horizontal direction and 2\% for the vertical direction.
In our view, it seems that such a measurement would be highly sensitive to outliers.
\citet{Blignaut2012a} primarily used a graphical analysis to illustrate the linearity of their device, though they did not specifically use the term linearity.
We think this is a good approach, but linearity could be further characterized statistically.

\subsubsection{Crosstalk}
In some of the earlier references to crosstalk (also known as ``cross-coupling''), little detail is provided on the precise calculation.
For example, the first formal measure of crosstalk in eye movements that we are aware of was by \citet{Merchant1967} who reported a crosstalk of 20\% maximum but provided no further details.
Similarly, \citet{Young1975} discussed crosstalk as a problem with \ac{EOG} without indicating the measurement method.

%To measure horizontal crosstalk due to vertical movement, \citet{Reulen1988} first plot the horizontal and vertical eye-movement signals against the vertical target positions.
%They perform a linear regression on the two eye-movement signals.
To describe the approach employed by \citet{Reulen1988} for the measurement of crosstalk, consider horizontal crosstalk as an example.
First, they plotted both the horizontal and vertical eye-movement signals against the vertical target signal.
Next, they performed a linear regression on both eye-movement signals.
This produced two slopes: one for horizontal eye position versus vertical target position, and one for vertical eye position versus vertical target position.
Horizontal crosstalk was then expressed as the ratio of the horizon\-tal-vertical slope to the vertical-vertical slope.
Analogous measurements were made for vertical crosstalk.
They found that this ratio was about 1/10 for their device, meaning an eye rotation of $10\degree{}$ in one direction caused $1\degree{}$ crosstalk in the orthogonal direction.
This approach assumed crosstalk was linear.

Working with a simulated model, \citet{Rigas2018} measured crosstalk as the absolute ratio (expressed as a percent) between the observed movement in one direction (horizontal or vertical) and the ground truth movements in the orthogonal direction.  
Their approach provided no information on the shape of the crosstalk effect (e.g., linear, quadratic, or cubic).

\subsubsection{Recalibration}
Generally, researchers use a \ac{MSC} at the start of each recording (though, there are efforts to make calibration-free eye-tracking devices \citep{Klefenz2010,Nagamatsu2009}).
Many researchers have started providing their own \acp{USC} after the \ac{MSC} and before task data is collected.
Typically, during a \ac{USC}, between 4 and 40 calibration targets are presented, distributed over some horizontal and vertical range.
However, some methods do not require calibration targets at all \citep{Takegami2002}.
\acp{USC} that do include targets determine when the subject is fixated on a target using one of three approaches \citep{Goldberg2003}: algorithm-controlled \citep{Akkil2014}, operator-controlled \citep{Nystrom2013}, or participant-controlled \citep{Blignaut2014,Hooge2018,Nystrom2013}.

After selecting samples during stable fixation, the actual recalibration of eye position takes place, and there are several methods for this.
One approach is to use a simple linear mapping, but this does not account for interactions between the two dimensions \citep{Blignaut2014b}.
More complex polynomial mappings can also be done \citep{Blignaut2014b,Kasprowski2014}, with the caveat that more calibration targets (and, thus, more time) are required to fit the polynomials.
Others create a 3D eye model to fit the measured eye movements \citep{Guestrin2006,Nagamatsu2008}.
Another approach is to perform Procrustes analysis \citep{Rosengren2019}, where the eye-position samples are translated, scaled, and rotated to fit the target positions.

Another issue is the time between the \ac{USC} and task-related data collection.
There is reason to believe that calibrations may deteriorate over time \citep{Hornof2002}, and therefore the shorter this interval, the better.
In the present study, our \ac{USC} was part of our experimental task, and so there was no delay between the collection of calibration data and the collection of the experimental data.

% Saccade (missing sharp edges in binocular)
% 2-3 catch-up saccades in one visualization
% Can see saccade in L/R but not with binocular

% Left Unfiltered | Left Filtered w/ SGolay | Binocular
% Large saccade
% Small catch-up saccade
% Goal: can still see details of saccades with filtered monocular data but not with binocular data

% {\color{red}
% Filtering signal:
% \begin{enumerate}
%     \item \citet{Kolarik2010} filtering reduces ``shake'' (precision)
%     \item \citet{Nystrom2013} filtering improves poor precision
%     \item \citet{Reingold2014} manufacturers are incentivized to filter their signals causing distorted velocity profiles, which affects the kinematics of saccadic eye movements, the ability to detect small saccades, and eye movements produced while looking at dynamic stimuli such as smooth pursuit
%     \item \citet{Hooge2018} some manufacturers apply filtering to decrease ``variable error'' (precision)
% \end{enumerate}}

\subsubsection{Filtering}
There is a general awareness of the potential influence of manufacturer-supplied filters of gaze position signals.
For example, in a study of spatial accuracy (``systematic error'') and spatial precision (``variable error''), \citet{Hooge2018} stated:
\begin{quote}
    ``The values for the decrease in the systematic error, and the increase in the variable error here presented, may be specific to the eye tracker, the experimental conditions and the populations we employed.
    Values may be different for different eye trackers (for example because some manufacturers apply filtering to the raw data to decrease the variable error), ...''
\end{quote}
Also, there is general knowledge about the potential negative effects of filtering on eye-movement signals.
For example, \citet{Reingold2014} made the following points:
\begin{quote}
    ``The emphasis on the tracking of ``stationary'' biological or artificial eyes as a primary method for eye tracker data quality evaluation has the unfortunate consequence of creating an incentive for manufacturers to produce systems that use heavy filtering (i.e., denoising algorithms) that, while making the eye look stable during fixations, severely distort the eye movement signal in terms of the velocity profile of the motion.
    Although appearing to improve static accuracy, it is often not appreciated that such filtering destroys important aspects of data quality including the temporal accuracy of identifying the beginning and end of fixations, the number of fixations detected (see Holmqvist et al., 2012), the kinematics of saccadic eye movements, the ability to detect small saccades, and eye movements produced while looking at dynamic stimuli (e.g., smooth pursuit).''
\end{quote}
For other comments on the negative effects of filtering, see \citet{Kolarik2010} and \citet{Nystrom2013}.
Therefore, we thought that it might be important to check for, and assess, any potential pre-filtering of our signals.
To this end, we performed a Fourier analysis of our various signals.
We propose that this sort of analysis should be considered as a part of any comprehensive assessment of eye-tracker quality.

\subsection{Present experimental plan}
In the present report, we applied new and potentially useful analyses to the characterization of the \ac{ET-HMD}.
Of course, we assessed spatial accuracy, spatial precision, temporal precision, linearity, and crosstalk.
% With the exception of crosstalk, all of our assessments were based on the analysis of a random saccade task. 
All of our assessments were based on the analysis of a random saccade task.
We introduced a method to remove the saccade latency for analysis of fixations during calibration, providing more data for analysis.
For spatial precision, we employed the non-parametric \ac{MAD} rather than the \ac{SD} so that our measure was robust to various underlying distributions.
In addition to a nominal sampling rate of 250~Hz, the \ac{ET-HMD} provides actual timestamps with nanosecond precision.
We characterized temporal precision in the present study as the \ac{SD} of \acp{ISI}.
Our user-supplied calibration was prepended to our task to minimize the time between recalibration and the recording of experimental data.
We presented a novel and sound ``binning'' method for the selection of samples within fixations to include in recalibration.
For linearity, we fit a line to the relationship between target position and eye position and described the degree of linearity with an \ac{R2} from a linear regression analysis.
Since an ideal system would have a slope of 1.0 for this fit, we compared the measured slope and its confidence limits to this ideal.
For our crosstalk analysis, we fit 4 multiple regression models (linear only, quadratic only, both linear and quadratic, and intercept only) and chose the best model using the Akaike information criterion~(AIC).
The \ac{ET-HMD} provides 3 signals (a left eye signal, a right eye signal, and a binocular signal) and all of these were compared on all quality metrics.
We also provided a Fourier analysis of the binocular and monocular signals and noted that the binocular signal is a substantially low-pass filtered version of the monocular signal.

% This report has the following structure:
% \autoref{sec:participants} describes our subjects,
% \autoref{sec:definitions} provides definitions and calculations for the data quality measures that we employ in this document,
% \autoref{sec:hardware} describes the eye-tracking hardware we use in this research,
% \autoref{sec:stimulus} discusses the stimulus and tasks utilized to accomplish this research,
% \autoref{sec:processing} details our data processing methods including saccade latency minimization and outlier filtration,
% \autoref{sec:recalibration} details our approach to recalibrating the gaze position signals,
% \autoref{sec:statistics} explains the statistical analyses we performed,
% \autoref{sec:results} contains our data quality results and statistical analyses,
% and \autoref{sec:discussion} discusses our findings and compares our approaches to the literature.

\section{Methods}
\label{sec:methods}

\subsection{Participants}
\label{sec:participants}
Twelve participants (9~males, 3~females, median age: 20, range: 19-66) willingly took part in this study.
%2~participants were faculty at Texas State University, 6~participants were college-aged students attending Texas State University, and the other 4~participants were engaging in a summer research experience for undergraduates~(REU) at Texas State University.
Five participants normally wore glasses but removed them for this study, three participants wore contact lenses during the study, and four required no vision correction.

\subsection{Definitions of data quality}
\label{sec:definitions}

\subsubsection{Spatial accuracy}
Following \citet{Holmqvist2012}, this is a measurement of the distance (in degrees of the visual angle) between the actual gaze position (where the subject is actually looking) and the measured gaze position (where the eye tracker reports the subject is looking) during stable fixation.
%For the present purposes, once a subject has fixated on a target, we treat the target position as the actual eye position.
For the present purposes, once a subject has fixated on a target, we assume the subject is actually looking at the target and that the target position can be treated as the actual eye position.
A separate measurement may be taken from the horizontal and vertical distances, or a single, combined measurement for both directions may be obtained (e.g., by using the Euclidean distance).
The dependent variable for statistical analyses of spatial accuracy is the mean accuracy per subject across fixations.

Consider a series of $n$ gaze samples recorded during a stable fixation.
Each sample includes a measured gaze position $(x^g_i,y^g_i)$ and a target position $(x^t_i,y^t_i)$, each in units of degrees of visual angle.
Note that since the target position remains constant during stable fixation, $(x^t_i,y^t_i)$ is the same for all $i$.
The spatial accuracy of the series of gaze samples is calculated using one of the following equations:
\begin{subequations}
\begin{align}
%&\text{Horizontal spatial accuracy}
\theta_H &= \frac{1}{n} \sum_{i=1}^n{|x^g_i - x^t_i|} \label{eqn:acc_h}\\
%&\text{Vertical spatial accuracy}
\theta_V &= \frac{1}{n} \sum_{i=1}^n{|y^g_i - y^t_i|} \label{eqn:acc_v}\\
%&\text{Combined spatial accuracy}
\theta_C &= \frac{1}{n} \sum_{i=1}^n{\sqrt{(x^g_i - x^t_i)^2 + (y^g_i - y^t_i)^2}} \label{eqn:acc_c}
\end{align}
\end{subequations}
where $\theta_H$ is horizontal accuracy, $\theta_V$ is vertical accuracy, and $\theta_C$ is combined accuracy (taking into account both horizontal and vertical components).

\subsubsection{Spatial precision}
Although some define spatial precision as the variation in the measured gaze position signal \citep{Blignaut2012}, as noted above, we prefer to think of both accuracy and precision as average-distance measures with different reference points.
For accuracy, the reference point is the target position.
For precision, the reference point is the central tendency of a set of sample positions.
As with spatial accuracy, spatial precision must be measured during stable fixation.
A separate measurement may be taken across the horizontal and vertical position signals, or a single, combined measurement may be obtained (e.g., by using the Euclidean distance to the central value).
The dependent variable for statistical analyses of spatial precision is the mean precision per subject across fixations.

Spatial precision is commonly measured in one of two ways, using either the \ac{SD} of sample positions or the \ac{RMS} (see Equations~2 and 3 in \citet{Holmqvist2012}).
Both of these approaches give a higher weight to larger deviations than to smaller ones due to the squaring of deviations, which is not necessarily desirable~\citep{Gorard2005}.
Therefore, we have chosen a spatial precision measurement based on the \ac{MAD} of sample positions.
This measure gives equal weight to deviations of all sizes and is robust to various shapes of underlying distributions (e.g., skewed or non-Gaussian).

% Given a series of measured gaze positions, $(\hat{x}_i, \hat{y}_i)$, $i = 1..n$, we denote the geometric median of the positions as the point $(\tilde{\hat{x}}, \tilde{\hat{y}})$.
% Then the spatial precision is calculated using one of the following equations, where $M(\hat{x})$ denotes $median(\hat{x}_1,\hat{x}_2,\dotsc,\hat{x}_n)$:
Consider a series of $n$ gaze samples recorded during a stable fixation.
Each sample includes a measured gaze position $(x^g_i,y^g_i)$.
We denote the geometric median of the sample positions as the point $(\tilde{x^g},\tilde{y^g})$ (computed using R package \texttt{Gmedian}, \citet{Cardot2017}).
The spatial precision of the series of gaze samples is calculated using one of the following equations, where $M(x)$ is short for $median(x)$:
\begin{subequations}
\begin{align}
%&\text{Horizontal spatial precision}
\text{MAD}_H &= M(|x^g_i - M(x^g)|),\ i = 1 \dotsc n \label{eqn:pre-h}\\
%&\text{Vertical spatial precision}
\text{MAD}_V &= M(|y^g_i - M(y^g)|),\ i = 1 \dotsc n \label{eqn:pre-v}\\
%&\text{Combined spatial precision}
\text{MAD}_C &= \sqrt{M(|x^g_i - \tilde{x^g}|)^2 + M(|y^g_i - \tilde{y^g}|)^2},\ i = 1 \dotsc n \label{eqn:pre-c}
\end{align}
\end{subequations}
where $\text{MAD}_H$ is horizontal precision, $\text{MAD}_V$ is vertical precision, and $\text{MAD}_C$ is combined precision (taking into account both horizontal and vertical components).

\subsubsection{Temporal precision}
\label{sec:definitions-temporal}
In one sense, temporal precision can be defined as the variability of \acp{ISI} (we call this sense the ``ISI sense'').
In another sense, as \citet{Holmqvist2012} have suggested, temporal precision can be thought of as the variability in ``system latency,'' or the difference between the time of the actual movement of the eye and the time reported by the eye tracker (we call this sense the ``Holmqvist sense'').
It is easy to see that temporal precision in the Holmqvist sense is nearly perfect with analog systems such as \ac{EOG}, infrared limbus tracking, and scleral search coil.
Although the \ac{ET-HMD} does produce timestamps for every video frame with nanosecond precision, there is some variability in the intersample intervals.
We have no ground-truth measurement of when eye movements are initiated, so we cannot measure temporal precision in the Holmqvist sense.
We can, and do, measure temporal precision in the ISI sense.

The nominal sampling rate of the \ac{ET-HMD} is 250~Hz, but the \ac{ET-HMD} does not always achieve precise 4~ms intervals between timestamps.\footnotemark\
The nanosecond timestamps were converted to milliseconds before calculating the \acp{ISI}.

\footnotetext{In e-mail correspondence with our former contacts at SMI, we asked, ``What is the general stability of the sampling rate of the eye tracker? Are there ever any dropped frames?''
One contact responded, ``It's possible, yes. Unlike some of our other products, the ET-HMD has to use the computer for image processing and eye tracking, so a resource problem could lead to dropped frames. I wouldn't expect this to be common unless there is a problem.''}

Consider a series of $n$ gaze samples.
Each sample includes a timestamp $t_i$ measured in milliseconds.
An \ac{ISI} is the difference between consecutive timestamps:
\begin{equation}
\Delta t_i = t_i - t_{i-1},\ i > 1
\label{eqn:isi}
\end{equation}
% We use the mean \ac{ISI}, $\mu_{\Delta t}$, to compute the temporal precision, which is simply the sample \ac{SD} of \acp{ISI} (note that there are $n-1$ \acp{ISI}, so we divide by $n-2$):
% \begin{equation}
% %&\text{Temporal precision}
% \sigma_T = \sqrt{\frac{1}{n-2} \sum_{i=2}^n{(\Delta t_i - \mu_{\Delta t})^2}}
% \label{eqn:pre-t}
% \end{equation}
We compute temporal precision as the sample \ac{SD} of \acp{ISI}.

\subsubsection{Linearity}
\label{sec:definitions-linearity}
When we discuss the term linearity, we are referring to the relationship between measured gaze position and actual gaze position.
There are at least two senses of the term linearity relevant here.
In one sense (hereafter referred to as ``linearity fit''), we want to know how well the data fit a line.
For this, we use the \ac{R2}.
In another sense (hereafter referred to as ``linearity slope''), we want to know how close the slope of the relationship is to an ideal slope of 1.0.
For this, we look at the measured slope and its 95\% confidence limits.
The dependent variable for statistical analyses of linearity slope is the slope estimate per subject across fixations.
The dependent variable for statistical analyses of linearity fit is the \ac{R2} per subject across fixations.

\begin{figure}
    %\centering
    \includegraphics[width=\linewidth,trim={0, 0, 0, 1.4cm},clip]{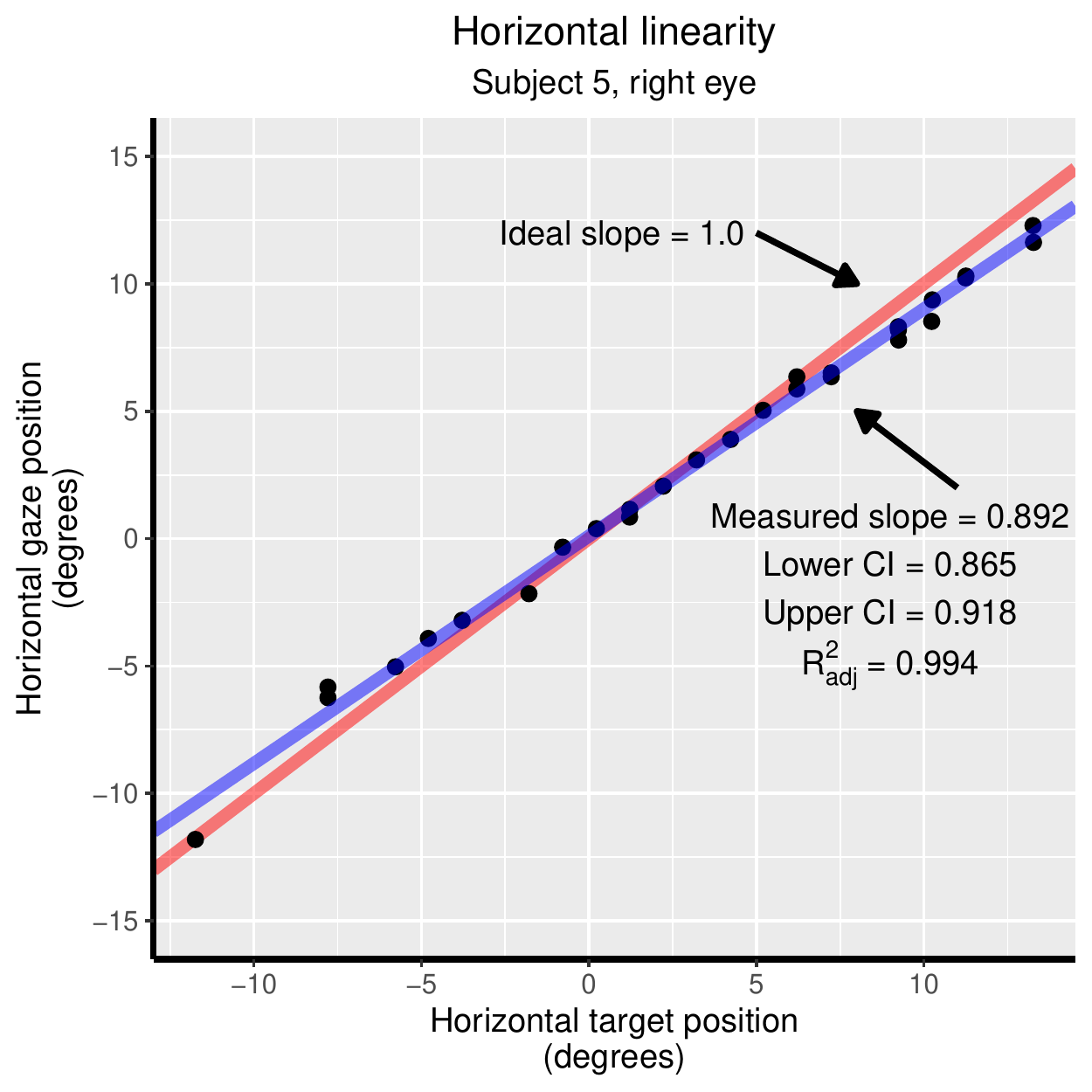}
    \caption{An example of linearity for illustrative purposes.}
    \label{fig:linearity-illustration}
\end{figure}

In \autoref{fig:linearity-illustration}, we present a linearity analysis for a single subject for illustrative purposes.
To assess horizontal linearity, we plot the horizontal gaze position versus horizontal target position.
Note that only one point is drawn per fixation target, because we perform our linearity regressions using the gaze centroid for each fixation instead of the individual gaze samples.
Both the linearity slope and linearity fit are shown in annotations on the figure.
In this case, the linearity slope is significantly different from the ideal slope of 1.0, because the 95\% confidence intervals do not include 1.0.

\subsubsection{Crosstalk}
\label{sec:definitions-crosstalk}
This is a measure of how much the rotation of the eye in one direction (horizontal or vertical) affects the measured gaze position in the orthogonal direction.
We refer to horizontal crosstalk from vertical movements as ``horizontal crosstalk,'' and to vertical crosstalk from horizontal movements as ``vertical crosstalk.''

The crosstalk measurements used in the past assume a linear relationship~\citep{Reulen1988,Rigas2018}, but we thought that it might be valuable to also consider the possibility of the crosstalk having a parabolic shape.
For data to assess crosstalk, we used fixation positions from our random saccade task, which sampled more-or-less the target movement range.
Our crosstalk assessments emerged from regression models, with the dependent measure as the offset of gaze position (in one direction, say horizontal) from target position (also horizontal in this case) and the independent variable was the target position in the orthogonal direction (vertical in this case).
All models fit an intercept.
Gaze offset is a similar concept to accuracy.
However, accuracy is a mean of the absolute value of distances, whereas offset is a mean value of distances which can be positive or negative.

For our crosstalk assessments, we used a step-wise approach and chose the best fitting model from the following 4 models: (1) a linear only fit, (2) a quadratic only fit, (3) both a linear and quadratic fit, and (4) an intercept only fit (neither linear nor quadratic).
%The criteria for the best model is based on a comparison between Akaike information criterion~(AIC).  The lower the AIC, the better the model.
When assessing the crosstalk for a single subject, the best model was decided by the \texttt{stepAIC} function of the R package \texttt{MASS} \citep{Venables2002}.
When assessing the crosstalk across subjects, the best model was decided by the \texttt{step} function of the R package \texttt{lmerTest} \citep{Kuznetsova2017}.

\begin{figure}
    %\centering
    \includegraphics[width=\linewidth,trim={0, 0, 0, 1.2cm},clip]{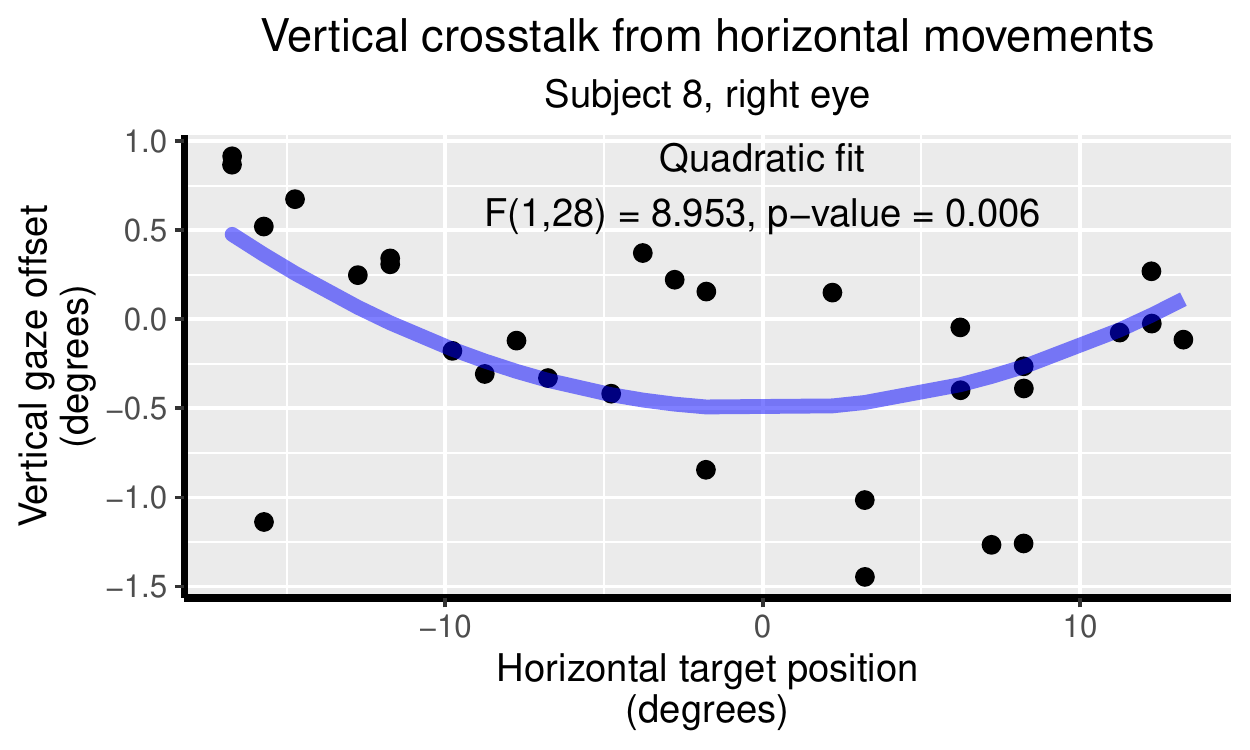}
    \caption{An example of crosstalk for illustrative purposes.}
    \label{fig:crosstalk-subject}
\end{figure}

In \autoref{fig:crosstalk-subject}, we present a crosstalk analysis for a single subject for illustrative purposes.
Note that only one point is drawn per fixation target, because we fit our crosstalk models using the gaze centroid for each fixation instead of the individual gaze samples.
To assess vertical crosstalk, we plotted the offset of the gaze centroid from the vertical target position versus horizontal target position.
In this case, the quadratic-only model was the best fit.

\subsection{Eye-tracking hardware}
\label{sec:hardware}

\subsubsection{\ac{ET-HMD}}
\label{sec:hardware-vive}
The \ac{ET-HMD} is a modified HTC Vive with an embedded SMI eye-tracking device.
The HTC Vive has dual Active-Matrix Organic Light-Emitting Diode (AMOLED) screens, each with a 3.6-inch diagonal and a resolution of 1080x1200 pixels (2160x1200 pixels combined).
%The factory specifications that we present below are taken from the brochure at \url{http://twittertechnews.com/wp-content/uploads/2016/09/smi_prod_eyetracking_hmd_HTC_Vive.pdf}.
The embedded eye-tracking device by SMI tracks both eyes simultaneously with a sampling rate of 250~Hz and has a typical spatial accuracy of $0.2\degree{}$.
\footnote{%
    \ac{ET-HMD} manufacturer-supplied specifications taken from \url{http://twittertechnews.com/wp-content/uploads/2016/09/smi_prod_eyetracking_hmd_HTC_Vive.pdf}.
}

%In addition to left and right eye signals, our device also provides a binocular signal.  There are no details available for exactly how this is computed.  We have been told by a representative of SMI and by others that the binocular trace provided by the Vive is smoothed (personnel communication).  Since this is undocumented, we can provide no details regarding this smoothing.

% \begin{center}
% \begin{tabular}{lll}
% Sampling rate & 250~Hz & (binocular)\\
% Spatial accuracy & $0.2\degree{}$ & (typical)
% \end{tabular}
% \end{center}

Each sample for each eye is associated with a 3-dimensional unit vector, $v = (v_x, v_y, v_z)$, that represents the direction the eye is looking.
Using standard trigonometry, we converted these 3-dimensional unit vectors into degrees of visual angle, $(x, y)$, with the help of MATLAB's \texttt{atan2d} function.
\begin{subequations}
\begin{align}
x &= \texttt{atan2d}(v_x, v_z)\\
y &= \texttt{atan2d}(v_y, v_z)
\end{align}
\end{subequations}
The \ac{ET-HMD} provides three of these direction vectors: left eye direction, right eye direction, and binocular direction (called the ``camera raycast'').
%We believe the binocular signal is smoothed in some way internally, but we do not know exactly what method and amount of smoothing has been performed.\footnotemark\
We have shown below (see Sect.~\ref{sec:fourier-result}) that the binocular signal is a low-pass filtered version of the monocular signal.
We adjusted the target position for monocular data following our observations shown in \autoref{fig:correct-target}, assuming an interpupillary distance of 62~mm.
No such adjustment was necessary for the binocular data.

%\footnotetext{In e-mail correspondence with our contacts at SMI, we asked, ``Is there any automatic smoothing of the eye signal, such as low or high pass filtering? If so, is there a way to disable this smoothing if we wanted to?''
%One contact responded, ``There's nothing optional -- there [is] some degree of built-in filtering [...] but this is fixed.''
%These are all the details we have about this question.}

\subsubsection{EyeLink~1000}
\label{sec:hardware-eyelink}
The EyeLink 1000 by SR Research is one of the leading eye-tracking devices currently on the market.
%The factory specifications that we present below are taken from the brochure at \url{http://sr-research.jp/support/EyeLink%201000%20User%20Manual%201.5.0.pdf}.
It tracks one eye with a sampling rate of 1000~Hz and has a typical spatial accuracy between $0.25\degree{}$ and $0.50\degree{}$.
\footnote{%
    EyeLink 1000 manufacturer-supplied specifications taken from    \url{http://sr-research.jp/support/EyeLink\%201000\%20User\%20Manual\%201.5.0.pdf}.
}

% \begin{center}
% \begin{tabular}{lll}
% Sampling rate & 1000~Hz & (monocular)\\
% Spatial accuracy & $0.25\degree{}$ to $0.50\degree{}$ & (typical)
% \end{tabular}
% \end{center}

The EyeLink data we used in the present research was collected by monocularly tracking the left eye.
We noticed the vertical data for our EyeLink was consistently $1.2\degree{}$ too low.
We do not know the source of this.
For the purposes of the comparisons performed in the present research, we simply added $1.2\degree{}$ in the vertical direction to every sample for all the EyeLink data.
There was no such offset in the horizontal direction.

\subsection{Description of the stimulus}
\label{sec:stimulus}
%The stimulus we employed for the Vive is a white, shaded sphere on a light gray background.
%It has a diameter of 100~mm and is positioned at a depth of 1000~mm, making it about $5.72\degree{}$ in diameter.
%We discuss the implications of such a large stimulus in \autoref{sec:discussion}.
We used Unity 2018.3.11f1 to create our stimulus for the \ac{ET-HMD}.
The stimulus was a small, solid-black sphere on a light gray background.
It had a diameter of $0.5\degree{}$ and was positioned at an apparent depth of 1000~mm (the actual target diameter was 8.72~mm, which was $0.5\degree{}$ at the chosen distance).

The primary eye movement task was a random saccade task.
Saccades could occur anywhere between $15\degree{}$ to the left and $15\degree{}$ to the right ($-15\degree{}$ to $+15\degree{}$ horizontally) and between $10\degree{}$ down and $10\degree{}$ up ($-10\degree{}$ to $+10\degree{}$ vertically), positioned relative to the nasal bridge, at a constant depth of one meter.
The position of each saccade target was determined randomly.
Thirty oblique saccades were performed, each with a minimum amplitude difference of $3\degree{}$ (radial) from the prior position.
The duration of each fixation was chosen from a uniform random distribution (min = 1~s, max = 1.5~s).

A calibration task (described in Sect.~\ref{sec:recalibration}) was prepended to the random saccade task to study the benefits of recalibration.

Headset position tracking was disabled so that, regardless of head orientation, the world center was always centered in the view.
As a result, the stimulus was always positioned correctly relative to the nasal bridge.

% A video of the task is available at \url{https://www.youtube.com/watch?v=n0GUu-WV6GA}.

For the EyeLink, the stimulus was a white ring on a dark background.
The ring had an inner diameter of about $0.5\degree{}$ and an outer diameter of about $1\degree{}$.
The recordings we used from the EyeLink were collected during a previous study using a different set of participants, but the eye movement task was also a random saccade task that was very similar to the one used for the \ac{ET-HMD} in this report.
The main differences are that 100~saccades were performed instead of just 30, and no user-supplied calibration task was prepended to the random saccade task.
Since the \ac{ET-HMD} task contained only 30~random saccades, we used only the first 30~saccades from the EyeLink task.

\subsection{Processing the gaze position signal}
\label{sec:processing}
\begin{figure}
	\centering
    \includegraphics[width=0.5\linewidth]{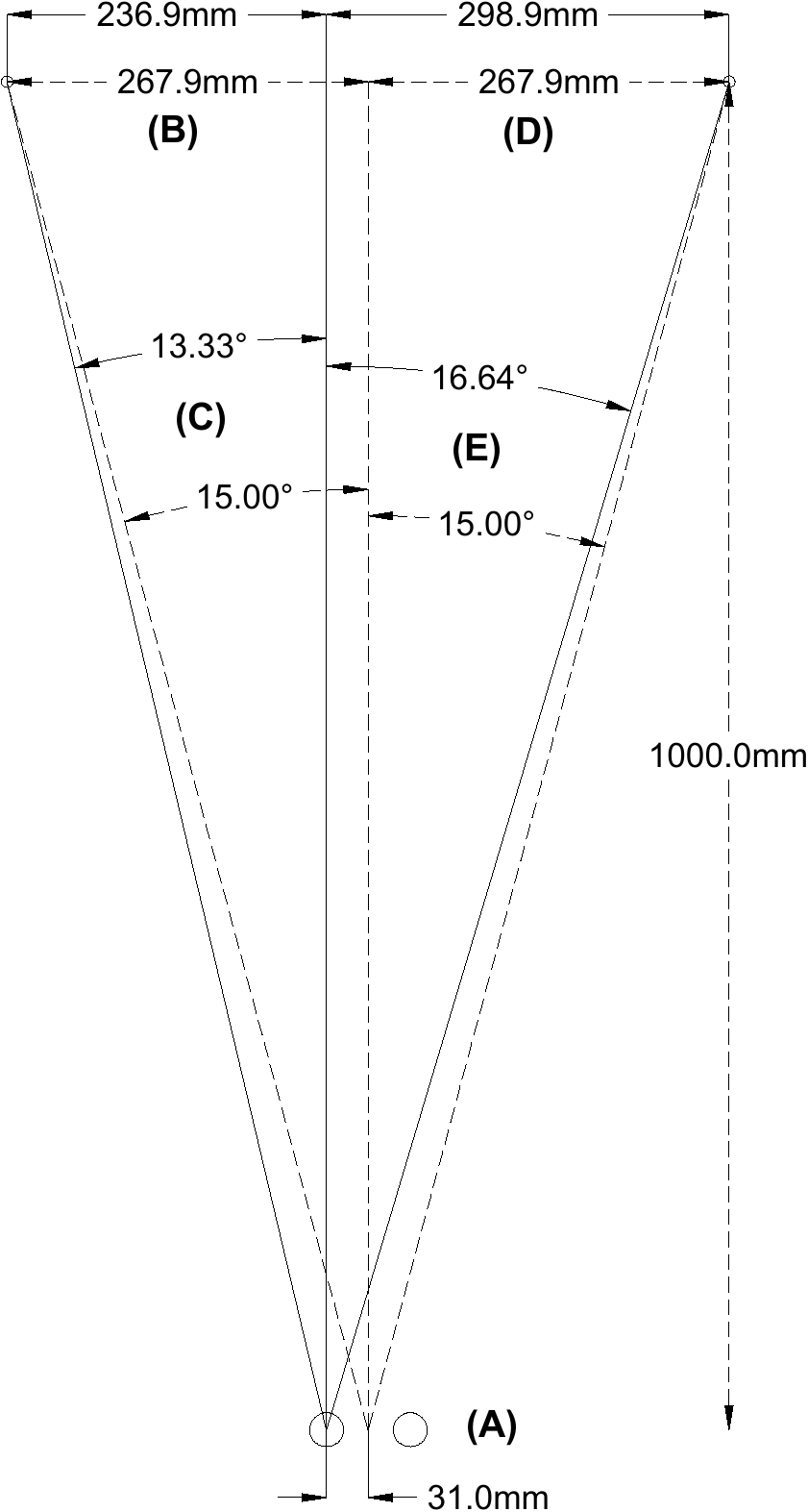}
	\caption{%
	%An illustration drawn to scale of the actual eye position compared to the stimulus position relative to the nasal bridge when the stimulus is 1000~mm away from the eyes.
	An illustration of the actual vs expected gaze angles for a stimulus positioned 1000~mm away from the eyes.
	We will consider the case of the left eye, but the same kind of problem occurs for the right eye, as well.
		(A) The two circles near the bottom of the figure represent the eyes, and the small circles near the top of the figure represent the stimulus.
		We assumed an interpupillary distance of 62~mm (31~mm from the nasal bridge to either eye).
		Dashed lines show the angle from the nasal bridge, while solid lines show the angle from the left eye.
		(B) When looking to the left, the stimulus is closer to the left eye (236.9~mm) than the nasal bridge (267.9~mm).
		(C) Therefore, a stimulus positioned $15\degree{}$ left of the nasal bridge is only $13.33\degree{}$ left of the left eye.
		(D) When looking to the right, the stimulus is further from the left eye (298.9~mm) than the nasal bridge (267.9~mm).
		(E) Therefore, a stimulus positioned $15\degree{}$ right of the nasal bridge is actually $16.64\degree{}$ right of the left eye.
		%Note that the opposite is true for the right eye (i.e., when looking at a target positioned $15\degree{}$ to the left of the nasal bridge, the right is actually looking $16.64\degree{}$ to the left; and when looking at a target positioned $15\degree{}$ to the right of the nasal bridge, the right eye is actually looking $13.33\degree{}$ to the right).
	}
% 		\textbf{(F)} It is worth mentioning that the stimulus we used has a diameter of 100~mm ($5.72\degree{}$ at 1000~mm distance), but a more commonly used target size would be between $0.2\degree{}$ and $1\degree{}$ in diameter.}
	\label{fig:correct-target}
\end{figure}

All targets were displayed relative to the nasal bridge, which is defined as the midpoint between the two eyes (``cyclopean eye'').
Thus, assuming an interpupillary distance of 62~mm, for a target positioned $-15\degree{}$ (horizontal), the left eye was actually viewing at $-13.3\degree{}$; and when the left eye was fixated on a target positioned $+15\degree{}$ (horizontal), the left eye was actually viewing at $+16.6\degree{}$.
We corrected for this disparity in the target position when evaluating each eye separately, but no correction was necessary when evaluating the binocular signal.
See \autoref{fig:correct-target} for an illustration of this phenomenon.
% TODO: Investigate the effect of this same correction on the EyeLink
% These corrections were not necessary for the data collected with the EyeLink.

\begin{figure}
	\centering
	\includegraphics[width=0.8\linewidth, trim={0cm 1cm 0.7cm 0.9cm}, clip]{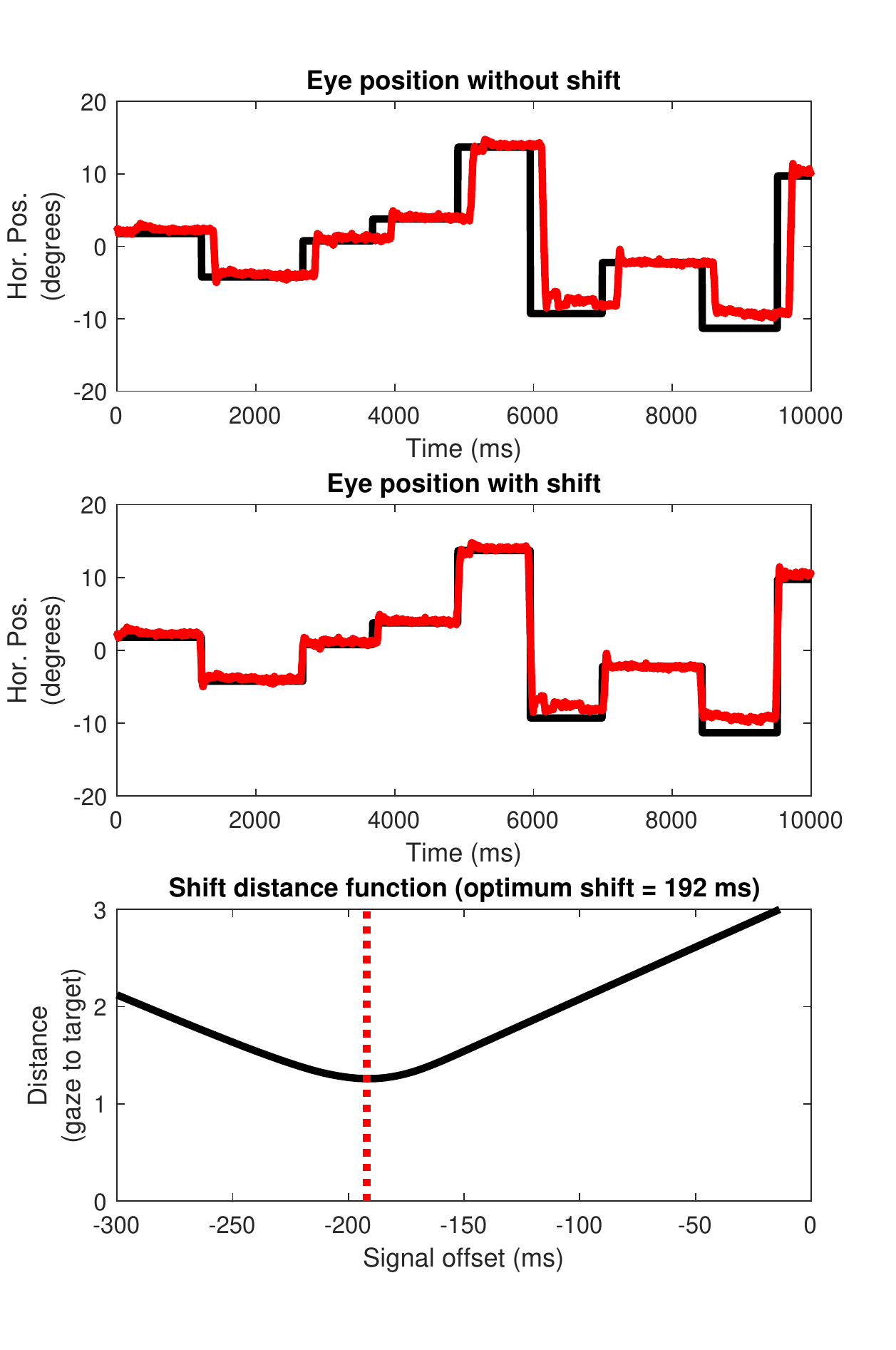}
	\caption{Illustration of our calculation of saccade latency.
		The top panel shows a typical saccade latency.
		The bottom panel illustrates the method we employed to determine the optimum shift to minimize the mean euclidean distance between the gaze position signal and the target position signal.
		We computed the mean euclidean distance between the two signals at various shifts and found the shift with the lowest mean distance (marked here with a red dashed line).
		The middle panel shows the data after the gaze position signal has been shifted by the optimum shift (in this case, 192~ms).
		}
	\label{fig:latency}
\end{figure}

The full gaze position signal contained various eye movements, including fixations, saccades, and blinks.
We wanted to measure data quality only when subjects were fixating.
Typically, the human reaction time to the saccadic movement of a target (saccade latency) is around 200~ms~\citep[p.~113]{Leigh2006}.
We first found the optimal temporal shift of the eye signal for each recording to align the eye and target movements as much as possible.
To obtain the best overall estimate of saccade latency, we calculated the mean Euclidean distance between the measured gaze position and the target position at shifts of 1~sample (4~ms, given the sampling rate of 250~Hz), from 1~sample to 200~samples (4~ms to 800~ms).
The shift resulting in the lowest mean Euclidean distance was chosen.
We illustrated this process in \autoref{fig:latency}.

% TODO: Change "Distance" axis to something like "Mean Euclidean Distance (gaze to target, deg)"

We neither manually nor algorithmically classified fixations and saccades.
Instead, we modeled the start of each fixation as the end of each target step, and the end of each fixation as the beginning of the next target step.
This was only reasonable because we had minimized the delay between the eye position data and the target position data, and also because all of the eye-tracking subjects were normal, healthy adults who followed the target very closely.

Each fixation duration was between 1000~ms and 1500~ms.
We discarded the first 400~ms of each fixation, since this was a time of some instability as the saccade, small corrective saccades, and post-saccadic oscillations transitioned into pure fixation.
The next 500~ms were employed for our data quality calculations.

\subsubsection{Removing outliers}
For these 500~ms, it was important to find and remove outliers prior to data quality calculation.
We used two screening steps for outlier removal (see \autoref{tbl:outliers} for outlier statistics):

\begin{enumerate}
\item
	We computed the first and third quartiles and the interquartile range (IQR) of the Euclidean distances between each measured gaze position and the centroid of the measured gaze.
	Any sample 1.5$\times$IQR below the first quartile or 1.5$\times$IQR above the third quartile was discarded (Tukey's fences~\citep{Tukey1977}).
\item
	We removed all samples that were more than $2\degree{}$ away from the centroid of the measured gaze (using the Euclidean distances calculated above).
\end{enumerate}

\begin{table}
	\centering
	%\captionsetup{font=normalsize}
	\caption{Outliers Removed per Subject}
% 		We used the \ac{ET-HMD}'s binocular position signal for these statistics, but outliers were filtered for each position signal.}
% 		Samples that were classified as an outlier in more than one filter step are grouped with the earliest one.}
	\label{tbl:outliers}
	\begin{threeparttable}
	\begin{tabular}{lcc}
\toprule

\multirow{2}[2]{*}{Subject} &
\multicolumn{2}{c}{\centering Outliers removed (\%)\tnote{\ssymbol{1}}} \\
\cmidrule{2-3}

{} & Screening step 1 & Screening step 2\tnote{\ssymbol{2}} \\

\midrule

1 & \mnsd{4.03}{6.73} & \mnsd{0.00}{0.00} \\
2 & \mnsd{4.58}{6.14} & \mnsd{0.00}{0.00} \\
3 & \mnsd{1.60}{2.42} & \mnsd{0.00}{0.00} \\
4 & \mnsd{2.75}{4.59} & \mnsd{0.00}{0.00} \\
5 & \mnsd{4.64}{7.02} & \mnsd{2.43}{13.33} \\
6 & \mnsd{3.89}{5.30} & \mnsd{0.00}{0.00} \\
7 & \mnsd{3.55}{6.59} & \mnsd{1.97}{10.81} \\
8 & \mnsd{5.86}{8.23} & \mnsd{1.70}{9.30} \\
9 & \mnsd{3.44}{4.91} & \mnsd{0.00}{0.00} \\
10 & \mnsd{3.87}{7.31} & \mnsd{2.40}{13.15} \\
11 & \mnsd{3.44}{4.89} & \mnsd{0.00}{0.00} \\
12 & \mnsd{2.83}{3.80} & \mnsd{0.00}{0.00} \\

\midrule

Mean\tnote{\ssymbol{3}} & \mnsd{3.71}{1.08} & \mnsd{0.71}{1.06} \\

\midrule
\end{tabular}

	\begin{tablenotes}[para,flushleft]
	    {\footnotesize
	        \item[\ssymbol{1}] Presented values are for the \ac{ET-HMD}'s binocular position signal.
	        
	        \item[\ssymbol{2}] These values do not include samples that were also classified as an outlier during step~1.
	        
	        \item[\ssymbol{3}] Mean and SD values across subjects.
	    }
% 		{\small %\footnotesize
% 		\ssymbol{1} Presented values are for the \ac{ET-HMD}'s binocular position signal.
		
% 		\ssymbol{2} These values do not include samples that were also classified as an outlier during step~1.
		
% 		\ssymbol{3} Mean and SD values across subjects.}
	\end{tablenotes}
	\end{threeparttable}
\end{table}

Finally, we calculated spatial accuracy and spatial precision for each fixation.
Linearity and crosstalk were measured across all fixations.
We repeated this process for all subjects and kept track of each fixation's data quality measures for later analysis.

% TODO: If we have to revise, think about binning only on the 500 ms we use for data quality assessment, rather than on the whole fixation.
\subsection{Description of recalibration}
\label{sec:recalibration}
We tested if our spatial accuracy measurements could be improved by employing a \ac{USC} of some kind.
This entailed prepending calibration targets prior to the random saccade task, and computing and applying recalibration coefficients.
Targets were displayed at one of 13~positions from a 13-point grid, ranging from top-left at ($-15\degree{}$, $+10\degree{}$) to the bottom right at ($+15\degree{}$, $-10\degree{}$).
The order of presentation of these positions was random.
Recalibration occurred after reducing saccade latency (see Sect.~\ref{sec:processing}).

\begin{figure}
\centering
\includegraphics[width=\linewidth, trim={3cm 0 2.5cm 1.68cm}, clip]{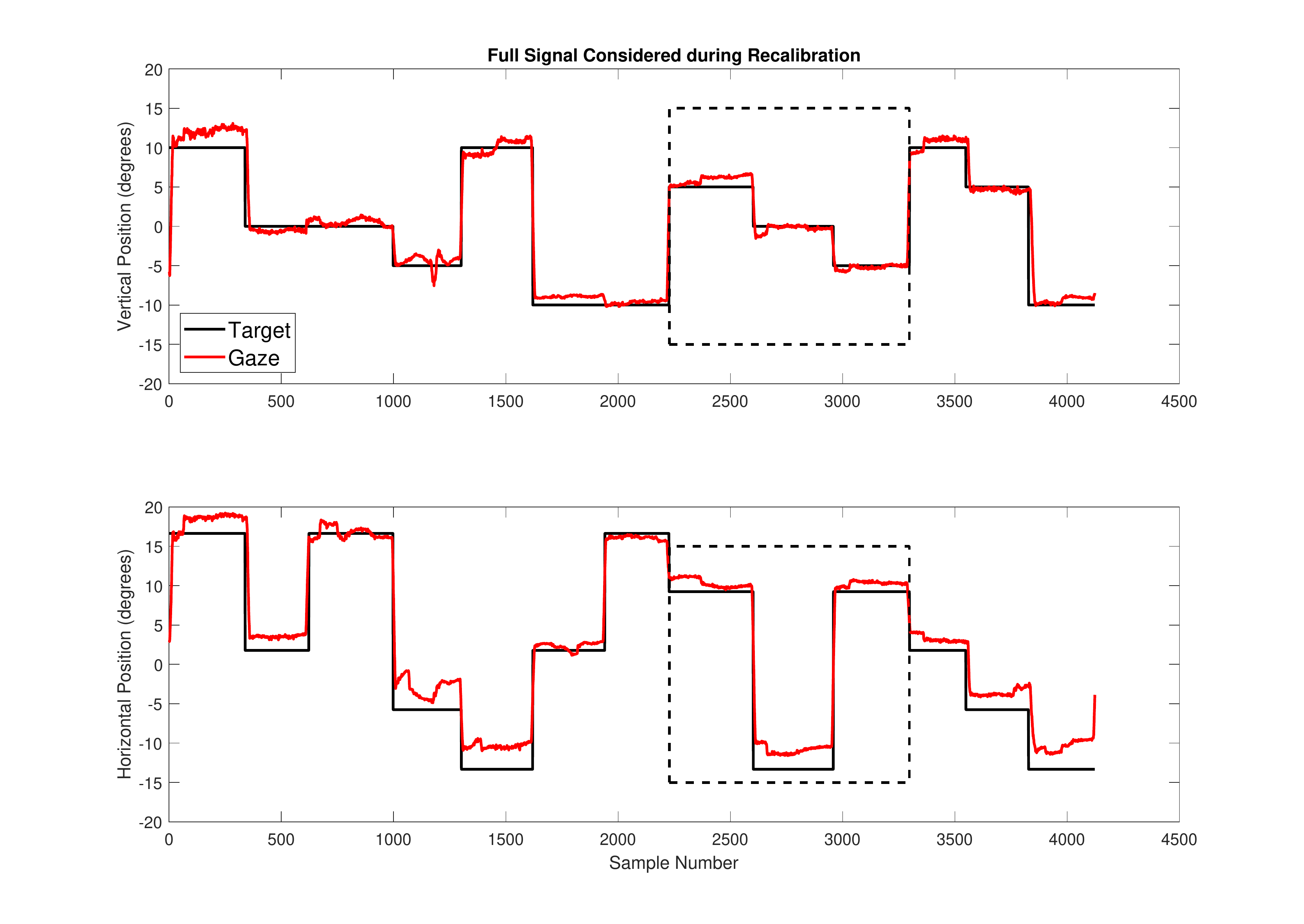}
\caption{The full signal used for recalibration for one of the subjects.
	The boxed portion of the signal is enlarged and used in \autoref{fig:recalibration-stable}.}
\label{fig:recalibration-bins}
\end{figure}

\begin{figure*}
\centering
\includegraphics[width=\linewidth, trim={3cm 0 2.5cm 1.69cm}, clip]{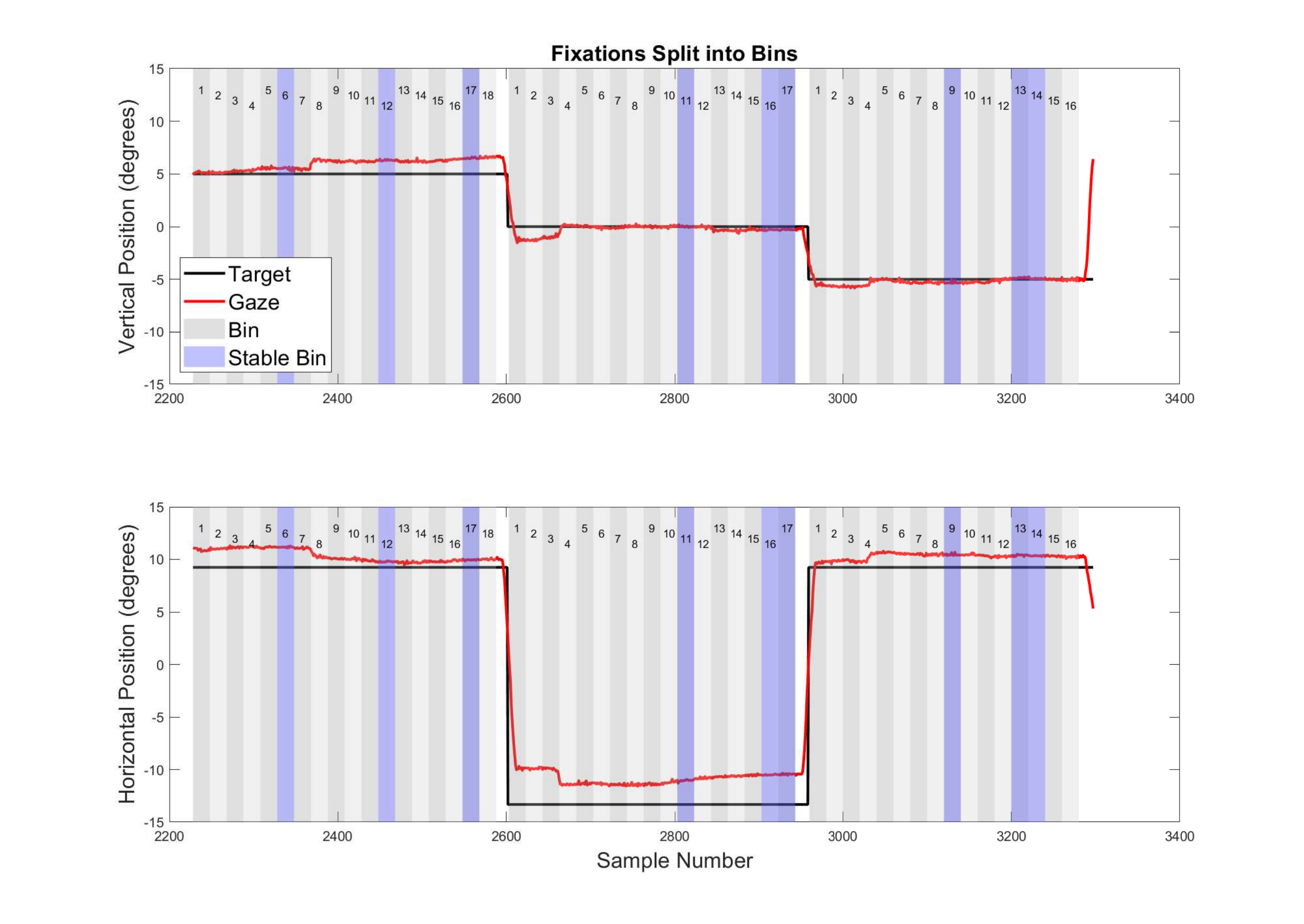}
\caption{%
    The most stable shared horizontal and vertical bins for a handful of fixations.
	Notice that each calibration target position has three ``stable bins.''
	The data in each of these stable bins are used for recalibration; the rest are ignored.
}
\label{fig:recalibration-stable}
\end{figure*}

We wanted to select samples from stable fixations on recalibration targets, with the idea that the eye is probably fixating on the target when it is moving least.
To identify stable fixations without classifying the signal, we split each fixation period (based on the target signal) into bins of 20~samples (80~ms).
Bins containing any invalid samples were dropped.
We computed the interquartile range (IQR) of the remaining bins as a measure of signal stability.
To combine the IQRs from the horizontal and vertical bins in a way that favors lower IQRs, we calculated a radial IQR as
\begin{equation}
IQR_{radial} = \sqrt{{IQR_{x}}^2 + {IQR_{y}}^2}.
\end{equation}
The three bins with the lowest radial IQR for each fixation were used for recalibration.
% In some cases, the most stable bins are during the short fixation period before a corrective saccade.
% We tried ignoring the first 500~ms of each fixation to prevent corrective saccades from being considered during the binning process, but this surprisingly led to worse recalibration performance.
We used the full fixation period for binning.
\autoref{fig:recalibration-bins} shows one of the signals obtained during the recalibration task.
\autoref{fig:recalibration-stable} shows the selected bins for the horizontal and vertical position signals.

% TODO: If we have to revise, fit the xlim to the data.

Finally, the horizontal and vertical gaze positions from the most stable bins were calibrated separately using the following equations from \citet{Kasprowski2014}:

\indent Linear:
\begin{subequations}
\begin{align}
x' &= A_{x}x + B_{x}y + C_{x}\\
y' &= A_{y}x + B_{y}y + C_{y}
\end{align}
\end{subequations}

\indent Quadratic:
\begin{subequations}
\begin{align}
x' &= A_{x}x^2 + B_{x}y^2 + C_{x}x + D_{x}y + E_{x}\\
y' &= A_{y}x^2 + B_{y}y^2 + C_{y}x + D_{y}y + E_{y}
\end{align}
\end{subequations}
where $A,B,C,D,E$ are the weights calculated by the regression procedure.
Note that all of these equations account for potential crosstalk between horizontal and vertical signals.

All of the data in this report were calibrated with a manu\-facturer-supplied calibration routine (for both the \ac{ET-HMD} and the EyeLink).
In what follows, we use ``NO-USC'' to mean no additional user-supplied calibration was applied beyond that provided by the device.
We refer to the linear calibration as USC-1 and the quadratic calibration as USC-2.
Each recalibration regression function was tested on the data obtained during the random saccade task.

\subsection{Statistical analyses}
\label{sec:statistics}
When evaluating the data quality across eyes within the \ac{ET-HMD} data, the statistical procedures will be most sensible if we imagine a particular dependent variable---say, for example, horizontal accuracy.
As noted above, each subject has one estimate of horizontal accuracy for each recording.
To compare horizontal accuracy across eyes, we have one fixed effect with three levels of "eye," and we have 12 subjects.
Because the subjects are the same across eyes, the "eye" effect is a repeated-measures (or within-subjects) effect.
This eye fixed effect was evaluated with a mixed model with subjects modeled as a random effect (R package \texttt{lmerTest}, \citet{Kuznetsova2017}).
Degrees of freedom were estimated using the Satterthwaite method (the default method for \texttt{lmerTest}).
A statistically significant F-value for the effect of "eye" was followed up with Dunnett-style contrasts with ``binocular'' as the reference level (R package \texttt{multcomp}, \citet{Hothorn2008}), and the Holm-Bonferroni method \citep{Holm1979} was used to control for multiple comparisons.
All p-values were two-tailed.

When evaluating the effect of calibration method on data quality within the \ac{ET-HMD} data, let's keep horizontal accuracy in mind as the dependent variable.
We had two statistical questions related to calibration: (1) Is calibration effective in improving data quality? (2) Are the two calibration methods different in effectiveness?
All of the calibration effects were tested using mixed models with subjects treated as a random effect (again using the \texttt{lmerTest} package \citep{Kuznetsova2017}).
We employed Helmert-style contrasts (again using the \texttt{multcomp} package \citep{Hothorn2008}) with a planned comparison approach.
Therefore, each statistical test for calibration gave us two p-values: one for the effect of calibration, and one for the difference between the two calibration methods.
The Holm-Bonferroni method \citep{Holm1979} was again employed to control for the presence of multiple tests.
All p-values were two-tailed.

To compare data quality from the \ac{ET-HMD} to the EyeLink~1000, we note that the data from these two devices come from completely independent subjects, and so a between-subjects design is appropriate.
We conducted several one-way analyses of means (R package \texttt{stats}, \citet{R2019}), not assuming equal variances.
(These are the exact statistical equivalent of a Welch's t-test.)
A statistically significant F-test indicates that the two devices have different horizontal accuracy (for example).
Inspection of the means indicated the direction of the effect.
All p-values were two-tailed.

Regarding linearity slopes, the above comparisons only test whether the slopes are significantly different between groups.
In order to test the significance of the slopes themselves, we also performed a one sample t-test comparing slopes to the ideal slope of 1.0.

% To investigate the effects of Eye and Calibration Method on the data quality of the \ac{ET-HMD}, we conducted several mixed model ANOVAs (R package \texttt{lmerTest}, \citet{Kuznetsova2017}).
% Degrees of freedom were estimated using the Satterthwaite method (the default method for \texttt{lmerTest}).
% For any significant effects, we followed up with post-hoc tests (R package \texttt{multcomp}, \citet{Hothorn2008}) to determine where the significance was.
% When Eye was the fixed effect, we used Dunnett-style contrasts (many-to-one) with binocular as the control.
% When Calibration Method was the fixed effect, we used Tukey-style contrasts (many-to-many).
% In both cases, the p-values were corrected using the Holm-Bonferonni method~\citep{Holm1979} to account for the multiple comparisons.

% To investigate the effects of Device on data quality between the \ac{ET-HMD} and the EyeLink, we conducted several one-way analyses of means (R package \texttt{stats}, \citet{R2019}), not assuming equal variances.
% Since the participants are different between devices, we did not need to treat Subject as a random effect.
% For any significant effects, we compared across-subjects means for each device to see which was better.

\subsubsection{Data transformations}
\label{sec:transformations}
Mathematical transformations were employed to enhance the normality of the underlying distributions of dependent measures prior to statistical analysis.  
Both spatial accuracy and spatial precision were, as a general matter, right-skewed, and a cube-root transformation improved normality for these measures. 
Linearity \ac{R2} are proportions, so a logit-transform was effective.
No transformation was necessary for linearity slope.

% TODO: Think about moving this e.g. after crosstalk in methods.
\subsection{Comparing left eye, right eye, and binocular signals using Fourier analysis}
\label{sec:fourier-method}
It was important to characterize the frequency content of our signals.
To this end, we employed a Fast Fourier Analysis (FFT).
For each of 12 subjects, we chose 3 stable fixation periods manually that were all 256 samples in length.
Data from each fixation period was mean-centered and detrended using a second-order polynomial fit.
Next, a Hann window was applied.
A FFT was performed on each fixation period using MATLAB FFT functions.
The single-sided magnitude spectra were extracted for each fixation period.
Average magnitude spectra were created by averaging across all fixation periods (M=3 per subject) and all subjects (N=12) for a total of 36 fixation periods.
This was done for the left eye, the right eye, and the binocular signal.
As per \citet{Hooge2018}, we also created a ``version'' signal which is the literal average of the left and right eye signals.
Since the version signal appeared even more similar to the binocular signal than either the left or right eye alone, and since the version signal should, in theory, also be a good approximation of a binocular signal, we also analyzed the version signal in the same way.
Examination of the magnitude spectrum of the version signal and the binocular signal suggested to us that the binocular signal may simply represent a low-pass filtered form of the version signal.
We wanted to characterize the filter which might have transformed the version signal into the binocular signal, and we posted a question on the website StackExchange\footnote{\url{https://dsp.stackexchange.com/questions/60098/reverse-engineering-a-digital-filter/60137}} where user MBaz posted the following method to obtain the filter.

The goal is to find a filter $h(t)$ such that:
\begin{equation}
    \quad\quad\quad y(t) = x(t) * h(t),
\end{equation}
where $x(t)$ is the version signal and $y(t)$ is the binocular signal.
To find the filter, note that:
\begin{equation}
    \quad\quad\quad Y(f) = X(f)H(f),
\end{equation}
where $f$ indicates the Fourier domain.
So,
\begin{equation}
    \quad\quad\quad h(t) = \text{Inverse FFT}\{Y(f)/X(f)\}.
\end{equation}

% \begin{quote}
%     ``You want a filter $h(t)$ such that
%     \begin{equation}
%         y(t) = x(t) * h(t),
%     \end{equation}
%     where $x(t)$ is [the version signal] and $y(t)$ is [the binocular signal].
    
%     To find the filter, note that
%     \begin{equation}
%         Y(f) = X(f)H(f),
%     \end{equation}
%     so
%     \begin{equation}
%         h(t) = \text{IFFT}\{Y(f)/X(f)\}.\text{''}
%     \end{equation}
% \end{quote}

We computed h(t) for each of the three 256-sample fixation periods from each subject, and then averaged all 36 h(t) filter estimates to get an average filter.
We then used the MATLAB tool \texttt{freqz} to plot the frequency response of the filter.  

\section{Results}
\label{sec:results}

Sect.~\ref{sec:results-vive} assesses the data quality of the \ac{ET-HMD} at different levels of Eye and Calibration Method.
Sect.~\ref{sec:results-device} assesses the data quality of the \ac{ET-HMD} compared to the EyeLink.
All of the raw (untransformed) means and \acp{SD} of all of our quantification results are presented in \autoref{tbl:summary-results-pm}, which will be further discussed as the results section progresses. 

\begin{table*}
	%\footnotesize
	\centering
	%\captionsetup{font=normalsize}
	\caption{Results for spatial accuracy, spatial precision, and linearity for both the \ac{ET-HMD} (with each recalibration method) and the EyeLink.
	Values are presented as the mean~$\pm$~1~\ac{SD} across subjects.}
% 	\textit{Dim} denotes the measurement dimension: horizontal~(H), vertical~(V), or combined~(C).
% 	\textit{Eye} denotes the gaze signal: left eye~(L), right eye~(R), or binocular~(B).}
	\label{tbl:summary-results-pm}
	\begin{threeparttable}
	\begin{tabular}{@{}cccP{2cm}P{2cm}P{2cm}P{2cm}@{}}
\toprule
 
\multirow{2}[2]{*}{Device} &
\multirow{2}[2]{*}{Dim\ssymbol{1}} &
\multirow{2}[2]{*}{Eye\ssymbol{2}} &
\multirow{2}[2]{2cm}{\centering Spatial accuracy} &
\multirow{2}[2]{2cm}{\centering Spatial precision}	&
\multicolumn{2}{c}{Linearity} \\
\cmidrule{6-7}

{} & {} & {} & {} & {} & Slope & \acs{R2} \\

\midrule

\parbox[t]{3mm}{\multirow{9}[3]{*}{\rotatebox[origin=c]{90}{ET-HMD}}} &
\multirow{3}{*}{H} &
L &
\mnsd{0.892}{0.366} &
\mnsd{0.102}{0.022} &
\mnsd{0.965}{0.047} &
\mnsd{0.993}{0.012} \\
{} &
{} &
R &
\mnsd{1.041}{0.236} &
\mnsd{0.084}{0.015} &
\mnsd{0.976}{0.033} &
\mnsd{0.998}{0.001} \\
{} &
{} &
B &
\mnsd{0.375}{0.173} &
\mnsd{0.052}{0.010} &
\mnsd{0.990}{0.037} &
\mnsd{0.998}{0.004} \\
\cmidrule{3-7}

{} &
\multirow{3}{*}{V} &
L &
\mnsd{0.690}{0.518} &
\mnsd{0.109}{0.044} &
\mnsd{0.992}{0.055} &
\mnsd{0.991}{0.012} \\
{} &
{} &
R &
\mnsd{0.528}{0.272} &
\mnsd{0.092}{0.023} &
\mnsd{0.979}{0.040} &
\mnsd{0.995}{0.006} \\
{} &
{} &
B &
\mnsd{0.474}{0.301} &
\mnsd{0.059}{0.012} &
\mnsd{1.003}{0.042} &
\mnsd{0.996}{0.005} \\
\cmidrule{3-7}

{} &
\multirow{3}{*}{C} &
L &
\mnsd{1.242}{0.651} &
\mnsd{0.173}{0.053} &
- &
- \\
{} &
{} &
R &
\mnsd{1.251}{0.312} &
\mnsd{0.148}{0.027} &
- &
- \\
{} &
{} &
B &
\mnsd{0.671}{0.338} &
\mnsd{0.107}{0.023} &
- &
- \\
\cmidrule{2-7}

\parbox[t]{3mm}{\multirow{9}[3]{*}{\rotatebox[origin=c]{90}{ET-HMD + USC-1}}} &
\multirow{3}{*}{H} &
L &
\mnsd{0.566}{0.379} &
\mnsd{0.105}{0.025} &
\mnsd{0.991}{0.029} &
\mnsd{0.993}{0.012} \\
{} &
{} &
R &
\mnsd{0.358}{0.138} &
\mnsd{0.086}{0.016} &
\mnsd{1.000}{0.018} &
\mnsd{0.998}{0.002} \\
{} &
{} &
B &
\mnsd{0.382}{0.264} &
\mnsd{0.054}{0.011} &
\mnsd{0.999}{0.020} &
\mnsd{0.996}{0.006} \\
\cmidrule{3-7}

{} &
\multirow{3}{*}{V} &
L &
\mnsd{0.418}{0.210} &
\mnsd{0.110}{0.038} &
\mnsd{1.005}{0.028} &
\mnsd{0.992}{0.011} \\
{} &
{} &
R &
\mnsd{0.376}{0.150} &
\mnsd{0.093}{0.023} &
\mnsd{0.996}{0.029} &
\mnsd{0.995}{0.005} \\
{} &
{} &
B &
\mnsd{0.300}{0.118} &
\mnsd{0.059}{0.011} &
\mnsd{1.005}{0.016} &
\mnsd{0.996}{0.004} \\
\cmidrule{3-7}

{} &
\multirow{3}{*}{C} &
L &
\mnsd{0.776}{0.467} &
\mnsd{0.176}{0.051} &
- &
- \\
{} &
{} &
R &
\mnsd{0.575}{0.193} &
\mnsd{0.151}{0.027} &
- &
- \\
{} &
{} &
B &
\mnsd{0.542}{0.295} &
\mnsd{0.111}{0.029} &
- &
- \\
\cmidrule{2-7}

\parbox[t]{3mm}{\multirow{9}[3]{*}{\rotatebox[origin=c]{90}{ET-HMD + USC-2}}} &
\multirow{3}{*}{H} &
L &
\mnsd{0.548}{0.359} &
\mnsd{0.105}{0.024} &
\mnsd{0.989}{0.022} &
\mnsd{0.992}{0.013} \\
{} &
{} &
R &
\mnsd{0.366}{0.128} &
\mnsd{0.086}{0.016} &
\mnsd{1.000}{0.019} &
\mnsd{0.997}{0.004} \\
{} &
{} &
B &
\mnsd{0.376}{0.239} &
\mnsd{0.053}{0.010} &
\mnsd{0.996}{0.016} &
\mnsd{0.996}{0.008} \\
\cmidrule{3-7}

{} &
\multirow{3}{*}{V} &
L &
\mnsd{0.450}{0.242} &
\mnsd{0.109}{0.038} &
\mnsd{1.005}{0.033} &
\mnsd{0.991}{0.011} \\
{} &
{} &
R &
\mnsd{0.384}{0.154} &
\mnsd{0.093}{0.024} &
\mnsd{0.995}{0.034} &
\mnsd{0.994}{0.005} \\
{} &
{} &
B &
\mnsd{0.300}{0.122} &
\mnsd{0.059}{0.011} &
\mnsd{1.004}{0.018} &
\mnsd{0.996}{0.004} \\
\cmidrule{3-7}

{} &
\multirow{3}{*}{C} &
L &
\mnsd{0.784}{0.470} &
\mnsd{0.176}{0.050} &
- &
- \\
{} &
{} &
R &
\mnsd{0.588}{0.200} &
\mnsd{0.151}{0.028} &
- &
- \\
{} &
{} &
B &
\mnsd{0.536}{0.274} &
\mnsd{0.109}{0.025} &
- &
- \\
\cmidrule{1-7}

\parbox[t]{3mm}{\multirow{3}{*}{\rotatebox[origin=c]{90}{EyeLink}}} &
H &
L &
\mnsd{0.677}{0.297} &
\mnsd{0.059}{0.023} &
\mnsd{1.009}{0.025} &
\mnsd{0.982}{0.044} \\
{} &
V &
L &
\mnsd{0.773}{0.480} &
\mnsd{0.074}{0.027} &
\mnsd{1.002}{0.068} &
\mnsd{0.932}{0.156} \\
{} &
C &
L &
\mnsd{1.137}{0.528} &
\mnsd{0.141}{0.066} &
- &
- \\

\bottomrule
\end{tabular}

	\begin{tablenotes}
	{\footnotesize
	    \item[\ssymbol{1}] Denotes the measurement dimension: horizontal~(H), vertical~(V), or combined~(C).
	    
	    \item[\ssymbol{2}] Denotes the measured gaze signal: left eye~(L), right eye~(R), or binocular~(B).
	}
	\end{tablenotes}
	\end{threeparttable}
\end{table*}

\subsection{Analysis of \ac{ET-HMD} data quality}
\label{sec:results-vive}
Our results with respect to the \ac{ET-HMD} will be focused on two questions: (1) How do the data from the left eye, the right and and the binocular signal compare on measures of data quality? (2) What is the impact of our two different methods of recalibration on data quality measures?

In Sect.~\ref{sec:eye-effect}, we separately analyze the effect of Eye at each level of Calibration Method (NO-USC, USC-1, and USC-2).
Similarly, in Sect.~\ref{sec:calibration-effect}, we separately analyze the effect of Calibration Method at each level of Eye (left, right, and binocular).

\subsubsection{Effect of eye on data quality}
\label{sec:eye-effect}

% Calibration Method == NO-USC
First, we will consider the data when Calibration Method is NO-USC.
The means and \acp{SD} for the uncalibrated and untransformed data for all quality measures is in the top panel of \autoref{tbl:summary-results-pm}, labeled ``ET-HMD.''
The means for accuracy and precision of the binocular signal were consistently superior to the left and right eye.
The linearity slopes for the binocular signal were closer to ideal than that for the left and right eye.
For linearity fit, the binocular signal was either superior to or tied with the best performance from the left or right eye.

Statistical tests for these comparison are in the section labelled ``NO-USC'' in \autoref{tbl:stats-eye}.
All of the F-values are statistically significant, indicating that the data quality depends on eye.
Seventeen of 20 tests comparing either the left eye or the right eye to the binocular signal were statistically significant, indicating the superiority of the binocular signal for data quality in uncalibrated data.

\begin{table*}
    \centering
    %\footnotesize
    %\captionsetup{font=normalsize}
    \caption{
        Analysis of the effect of eye on data quality at different levels of calibration method.}
        % The F-statistics from the mixed model ANOVAs are presented.
        % Contrasts are also presented when an ANOVA is significant.}
        % ``L - B'' indicates the contrast between the left and binocular signals, and ``R - B'' indicates the contrast between the right and binocular signals.}
    \label{tbl:stats-eye}
    \begin{threeparttable}
    \begin{tabular}{@{}cccccccccccc@{}}
\toprule

\multirow{2}[2]{2cm}{\centering Calibration method} &
\multicolumn{2}{c}{\multirow{2}[2]{2cm}{\centering Dependent variable}} &
\multicolumn{2}{c}{DF\ssymbol{1}} &
\multirow{2}[2]{*}{F\ssymbol{2}} &
\multirow{2}[2]{*}{p} &
\multirow{2}[2]{*}{Sig.} &
\multicolumn{2}{c}{L - B\ssymbol{3}} &
\multicolumn{2}{c}{R - B\ssymbol{4}} \\
\cmidrule(lr){4-5}\cmidrule(lr){9-10}\cmidrule(lr){11-12}

{} & {} & {} & Num & Den & {} & {} & {} & Est. & p & Est. & p \\

\midrule

\parbox[t]{3mm}{\multirow{10}[5]{*}{\rotatebox[origin=c]{90}{NO-USC}}} &
\multirow{3}{*}{Accuracy} &
H &
2 & 22.00 & 35.93 & <.001 & \checkmark & 0.24 & <.001 & 0.30 & <.001 \\
{} &
{} &
V &
2 & 22.00 & 7.25 & .004 & \checkmark & 0.09 & <.001 & 0.03 & .173 \\
{} &
{} &
C &
2 & 22.00 & 28.69 & <.001 & \checkmark & 0.19 & <.001 & 0.21 & <.001 \\
\cmidrule{3-12}
{} &
\multirow{3}{*}{Precision} &
H &
2 & 22.00 & 67.47 & <.001 & \checkmark & 0.09 & <.001 & 0.06 & <.001 \\
{} &
{} &
V &
2 & 22.00 & 58.69 & <.001 & \checkmark & 0.09 & <.001 & 0.06 & <.001 \\
{} &
{} &
C &
2 & 22.00 & 45.57 & <.001 & \checkmark & 0.08 & <.001 & 0.05 & <.001 \\
\cmidrule{3-12}
{} &
\multirow{2}{*}{Lin. slope} &
H &
2 & 22.00 & 4.40 & .025 & \checkmark & -0.02 & .006 & -0.01 & .109 \\
{} &
{} &
V &
2 & 22.00 & 3.58 & .045 & \checkmark & -0.01 & .209 & -0.02 & .015 \\
\cmidrule{3-12}
{} &
\multirow{2}{*}{Lin. fit} &
H &
2 & 22.00 & 11.95 & <.001 & \checkmark & -0.90 & <.001 & -0.31 & .093 \\
{} &
{} &
V &
2 & 22.00 & 19.31 & <.001 & \checkmark & -0.69 & <.001 & -0.23 & .039 \\
\midrule

\parbox[t]{3mm}{\multirow{10}[5]{*}{\rotatebox[origin=c]{90}{USC-1}}} &
\multirow{3}{*}{Accuracy} &
H &
2 & 22.00 & 7.24 & .004 & \checkmark & 0.10 & .002 & 0.00 & .883 \\
{} &
{} &
V &
2 & 22.00 & 22.47 & <.001 & \checkmark & 0.07 & <.001 & 0.05 & <.001 \\
{} &
{} &
C &
2 & 22.00 & 10.04 & <.001 & \checkmark & 0.10 & <.001 & 0.03 & .216 \\
\cmidrule{3-12}
{} &
\multirow{3}{*}{Precision} &
H &
2 & 22.00 & 63.03 & <.001 & \checkmark & 0.09 & <.001 & 0.06 & <.001 \\
{} &
{} &
V &
2 & 22.00 & 69.59 & <.001 & \checkmark & 0.09 & <.001 & 0.06 & <.001 \\
{} &
{} &
C &
2 & 22.00 & 42.38 & <.001 & \checkmark & 0.08 & <.001 & 0.05 & <.001 \\
\cmidrule{3-12}
{} &
\multirow{2}{*}{Lin. slope} &
H &
2 & 22.00 & 0.95 & .403 & {} & {} & {} & {} & {} \\
{} &
{} &
V &
2 & 22.00 & 0.60 & .560 & {} & {} & {} & {} & {} \\
\cmidrule{3-12}
{} &
\multirow{2}{*}{Lin. fit} &
H &
2 & 22.00 & 8.06 & .002 & \checkmark & -0.85 & <.001 & -0.21 & .348 \\
{} &
{} &
V &
2 & 22.00 & 11.62 & <.001 & \checkmark & -0.61 & <.001 & -0.39 & .003 \\
\midrule

\parbox[t]{3mm}{\multirow{10}[5]{*}{\rotatebox[origin=c]{90}{USC-2}}} &
\multirow{3}{*}{Accuracy} &
H &
2 & 22.00 & 6.37 & .007 & \checkmark & 0.10 & .002 & 0.01 & .717 \\
{} &
{} &
V &
2 & 22.00 & 15.04 & <.001 & \checkmark & 0.09 & <.001 & 0.06 & <.001 \\
{} &
{} &
C &
2 & 22.00 & 12.19 & <.001 & \checkmark & 0.10 & <.001 & 0.03 & .105 \\
\cmidrule{3-12}
{} &
\multirow{3}{*}{Precision} &
H &
2 & 22.00 & 72.48 & <.001 & \checkmark & 0.09 & <.001 & 0.07 & <.001 \\
{} &
{} &
V &
2 & 22.00 & 69.90 & <.001 & \checkmark & 0.09 & <.001 & 0.06 & <.001 \\
{} &
{} &
C &
2 & 22.00 & 50.31 & <.001 & \checkmark & 0.08 & <.001 & 0.06 & <.001 \\
\cmidrule{3-12}
{} &
\multirow{2}{*}{Lin. slope} &
H &
2 & 22.00 & 1.66 & .213 & {} & {} & {} & {} & {} \\
{} &
{} &
V &
2 & 22.00 & 0.67 & .522 & {} & {} & {} & {} & {} \\
\cmidrule{3-12}
{} &
\multirow{2}{*}{Lin. fit} &
H &
2 & 22.00 & 7.84 & .003 & \checkmark & -0.89 & <.001 & -0.17 & .490 \\
{} &
{} &
V &
2 & 22.00 & 16.26 & <.001 & \checkmark & -0.67 & <.001 & -0.38 & .001 \\

\bottomrule
\end{tabular}

    \begin{tablenotes}
    {\footnotesize
        \item[\ssymbol{1}] Numerator and denominator degrees of freedom, estimated using the Satterthwaite method.
        
        \item[\ssymbol{2}] F-values from the mixed model ANOVAs.
    
        \item[\ssymbol{3}] Contrasts between the left and binocular signals.
        Omitted when the ANOVA is not significant.
        A positive estimate indicates the value from the left eye is higher (but not necessarily \textit{better}).
        
        \item[\ssymbol{4}] Contrasts between the right and binocular signals.
        Omitted when the ANOVA is not significant.
        A positive estimate indicates the value from the right eye is higher (but not necessarily \textit{better}).
    }
    \end{tablenotes}
    \end{threeparttable}
\end{table*}

% Calibration Method == USC-1
Next, we will consider the data after applying the USC-1 method of recalibration.
The means and SDs for the untransformed quality measures after USC-1 recalibration are in the second panel of \autoref{tbl:summary-results-pm}, labeled ``ET-HMD + USC-1.''
The means for vertical and combined spatial accuracy of the binocular signal are better than the left and right eye.
The means for spatial precision of the binocular signal are better than the left and right eye.
The linearity slopes are not much different between the binocular, left, and right eye signals.
For linearity fit, the binocular signal is either superior to or tied with the best performance from the left and right eye.

Statistical tests for these comparisons are in the section labeled ``USC-1'' in \autoref{tbl:stats-eye}.
The F-values for accuracy, precision, and linearity fit are all statistically significant, indicating that the data quality depends on eye.
Thirteen of 16 tests comparing either the left eye or the right eye to the binocular signal were statistically significant, indicating the superiority of the binocular signal for data quality even after recalibration with USC-1.

% Calibration Method == USC-2
Lastly, we will consider the data after applying the USC-2 method of recalibration.
The means and SDs for the untransformed quality measures after USC-2 recalibration are in the third panel of \autoref{tbl:summary-results-pm}, labeled ``ET-HMD + USC-2.''
Again, the means for vertical and combined spatial accuracy of the binocular signal are better than the left and right eye.
The means for spatial precision of the binocular signal are better than the left and right eye.
The linearity slopes are again not much different between the binocular, left, and right eye signals.
For linearity fit, the binocular signal is only better in the vertical direction.

Statistical tests for these comparisons are in the section labeled ``USC-2'' in \autoref{tbl:stats-eye}.
As with USC-1, the F-values for accuracy, precision, and linearity fit are all statistically significant, indicating that the data quality depends on eye.
Again, 13 of 16 post-hoc tests were statistically significant, indicating the superiority of the binocular signal for data quality even after recalibration with USC-2.

In \autoref{tbl:stats-linearity-slopes}, we present the results of the t-tests comparing each linearity slope to the ideal slope of 1.0.
As a general matter, the empirical slopes for linearity were not statistically different from the ideal slope.
However, for the horizontal slopes of the uncalibrated left and right eyes, the slopes were significantly lower than the ideal.

\begin{table*}
    \centering
    %\captionsetup{font=normalsize}
    \caption{
        Results from the t-tests comparing linearity slopes to the ideal of 1.0.}
        % A significant result indicates that the linearity slope is significantly different from the ideal slope.}
    \label{tbl:stats-linearity-slopes}
    \begin{threeparttable}
    \begin{tabular}{@{}cccccccccc@{}}
\toprule

\multirow{2}[2]{*}{Device} &
\multirow{2}[2]{*}{Eye} &
\multirow{2}[2]{*}{Dim} &
\multirow{2}[2]{*}{DF} &
\multirow{2}[2]{*}{t} &
\multirow{2}[2]{*}{p} &
\multirow{2}[2]{*}{Sig.{\ssymbol{1}}} &
\multirow{2}[2]{2cm}{\centering Slope estimate} &
\multicolumn{2}{c}{95\% conf. int.} \\
\cmidrule{9-10}

{} & {} & {} & {} & {} & {} & {} & {} & Lower & Upper \\

\midrule

\multirow{6}[4]{*}{ET-HMD} &
\multirow{2}{*}{L} &
H &
11 & -2.60 & .025 & \checkmark & 0.965 & 0.935 & 0.995 \\
{} &
{} &
V &
11 & -0.52 & .613 & {} & 0.992 & 0.957 & 1.027 \\
\cmidrule{3-10}

{} &
\multirow{2}{*}{R} &
H &
11 & -2.50 & .029 & \checkmark & 0.976 & 0.956 & 0.997 \\
{} &
{} &
V &
11 & -1.82 & .096 & {} & 0.979 & 0.953 & 1.004 \\
\cmidrule{3-10}

{} &
\multirow{2}{*}{B} &
H &
11 & -0.94 & .366 & {} & 0.990 & 0.966 & 1.014 \\
{} &
{} &
V &
11 & 0.27 & .796 & {} & 1.003 & 0.977 & 1.030 \\
\cmidrule{2-10}

\multirow{6}[4]{2cm}{\centering ET-HMD + USC-1} &
\multirow{2}{*}{L} &
H &
11 & -1.11 & .291 & {} & 0.991 & 0.972 & 1.009 \\
{} &
{} &
V &
11 & 0.60 & .561 & {} & 1.005 & 0.987 & 1.023 \\
\cmidrule{3-10}

{} &
\multirow{2}{*}{R} &
H &
11 & 0.05 & .961 & {} & 1.000 & 0.989 & 1.012 \\
{} &
{} &
V &
11 & -0.43 & .678 & {} & 0.996 & 0.978 & 1.015 \\
\cmidrule{3-10}

{} &
\multirow{2}{*}{B} &
H &
11 & -0.14 & .888 & {} & 0.999 & 0.986 & 1.012 \\
{} &
{} &
V &
11 & 1.06 & .311 & {} & 1.005 & 0.995 & 1.015 \\
\cmidrule{2-10}

\multirow{6}[4]{2cm}{\centering ET-HMD + USC-2} &
\multirow{2}{*}{L} &
H &
11 & -1.70 & .118 & {} & 0.989 & 0.975 & 1.003 \\
{} &
{} &
V &
11 & 0.54 & .602 & {} & 1.005 & 0.984 & 1.026 \\
\cmidrule{3-10}

{} &
\multirow{2}{*}{R} &
H &
11 & 0.00 & 1 & {} & 1.000 & 0.988 & 1.012 \\
{} &
{} &
V &
11 & -0.50 & .630 & {} & 0.995 & 0.973 & 1.017 \\
\cmidrule{3-10}

{} &
\multirow{2}{*}{B} &
H &
11 & -0.96 & .360 & {} & 0.996 & 0.985 & 1.006 \\
{} &
{} &
V &
11 & 0.85 & .412 & {} & 1.004 & 0.993 & 1.016 \\
\midrule

\multirow{2}{*}{EyeLink} &
\multirow{2}{*}{L} &
H &
9 & 1.08 & .309 & {} & 1.009 & 0.991 & 1.027 \\
{} &
{} &
V &
9 & 0.09 & .927 & {} & 1.002 & 0.954 & 1.050 \\

\bottomrule
\end{tabular}

    \begin{tablenotes}
    {\footnotesize
        \item[\ssymbol{1}] A significant result indicates that the linearity slope is significantly different from the ideal slope of 1.0.}
    \end{tablenotes}
    \end{threeparttable}
\end{table*}

Based on these results, if one had to choose only one signal based on accuracy, precision, and linearity, the binocular signal would have to be that choice.

\subsubsection{Effect of calibration method on data quality}
\label{sec:calibration-effect}
\begin{figure*}
    \centering
    \includegraphics[width=\linewidth,trim={0, 0, 0, 0.7cm},clip]{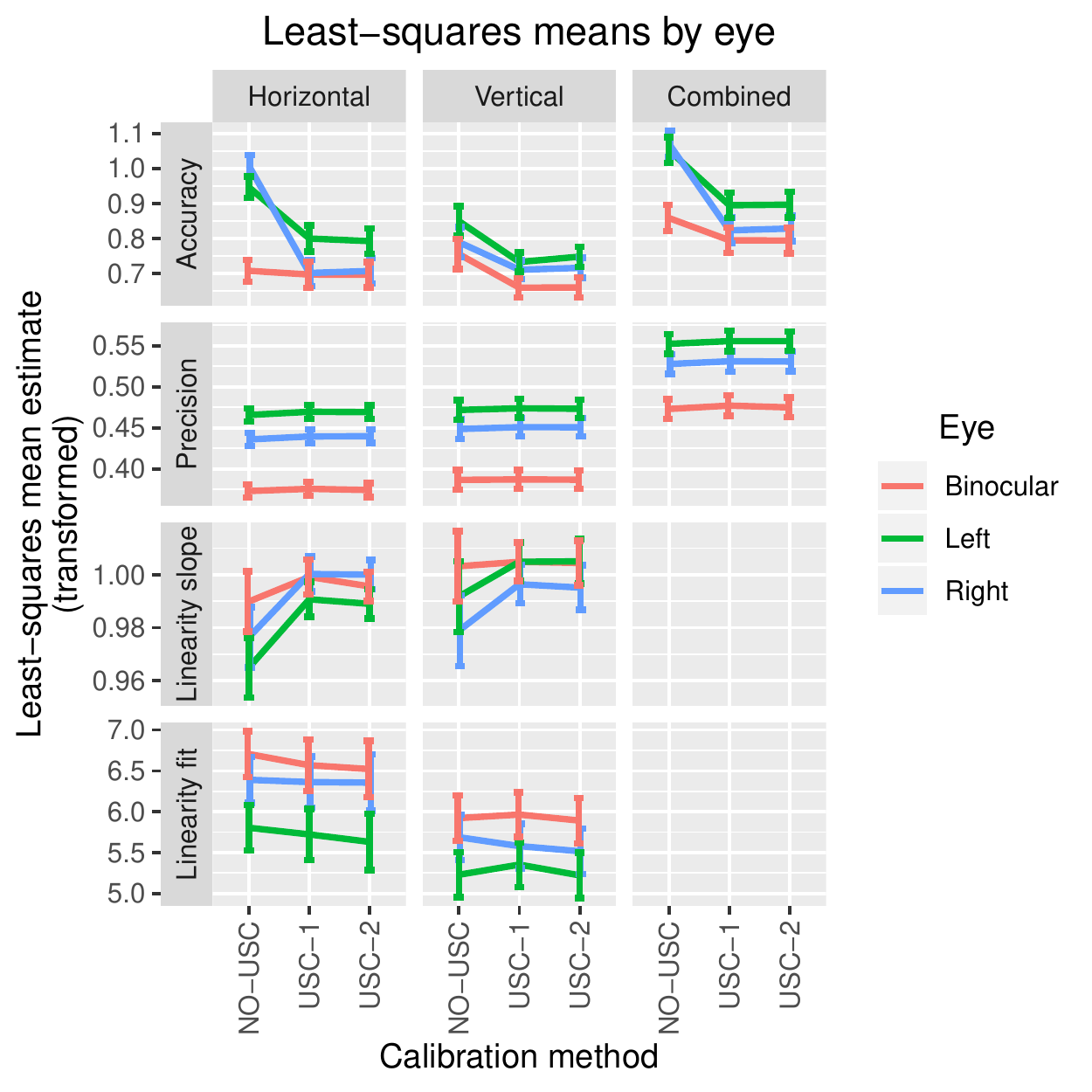}
    \caption{
        Least-squares means of data quality measures across eyes for each calibration method and dimension.}
    \label{fig:lsmeans}
\end{figure*}

The means and SDs for the untransformed data are in \autoref{tbl:summary-results-pm}, and the statistical analysis of the effect of calibration is presented in \autoref{tbl:stats-calibration}.
We also present a visualization of the comparisons in \autoref{fig:lsmeans}.

\begin{table*}
    \centering
    %\footnotesize
    %\captionsetup{font=normalsize}
    \caption{
        Analysis of the effect of calibration method on data quality for each eye.}
    \label{tbl:stats-calibration}
    \begin{threeparttable}
    \begin{tabular}{@{}ccccccccccccc@{}}
\toprule

\multirow{2}[2]{*}{Eye} &
\multicolumn{2}{c}{\multirow{2}[2]{2cm}{\centering Dependent variable}} &
\multicolumn{5}{c}{NO-USC vs mean(USC-1 and USC-2)} &
\multicolumn{5}{c}{USC-1 vs USC-2} \\
\cmidrule(lr){4-8}\cmidrule(lr){9-13}

{} & {} & {} & Est. & Std. Err. & z\ssymbol{1} & p & Sig. & Est. & Std. Err. & z\ssymbol{1} & p & Sig. \\

\midrule

\parbox[t]{3mm}{\multirow{10}[5]{*}{\rotatebox[origin=c]{90}{Left}}} &
\multirow{3}{*}{Accuracy} &
H &
0.15 & 0.03 & 4.71 & <.001 & \checkmark & 0.01 & 0.04 & 0.20 & .841 & {} \\
{} &
{} &
V &
0.11 & 0.03 & 4.04 & <.001 & \checkmark & -0.02 & 0.03 & -0.49 & .626 & {} \\
{} &
{} &
C &
0.16 & 0.03 & 6.12 & <.001 & \checkmark & -0.00 & 0.03 & -0.06 & .953 & {} \\
\cmidrule{3-13}
{} &
\multirow{3}{*}{Precision} &
H &
-0.00 & 0.00 & -2.41 & .031 & \checkmark & 0.00 & 0.00 & 0.27 & .785 & {} \\
{} &
{} &
V &
-0.00 & 0.00 & -0.68 & .993 & {} & 0.00 & 0.00 & 0.19 & .993 & {} \\
{} &
{} &
C &
-0.00 & 0.00 & -2.20 & .055 & {} & 0.00 & 0.00 & 0.02 & .982 & {} \\
\cmidrule{3-13}
{} &
\multirow{2}{*}{Lin. slope} &
H &
-0.02 & 0.01 & -3.28 & .002 & \checkmark & 0.00 & 0.01 & 0.19 & .846 & {} \\
{} &
{} &
V &
-0.01 & 0.01 & -1.12 & .522 & {} & -0.00 & 0.01 & -0.01 & .990 & {} \\
\cmidrule{3-13}
{} &
\multirow{2}{*}{Lin. fit} &
H &
0.13 & 0.06 & 1.99 & .094 & {} & 0.09 & 0.07 & 1.24 & .215 & {} \\
{} &
{} &
V &
-0.06 & 0.07 & -0.84 & .401 & {} & 0.13 & 0.08 & 1.64 & .201 & {} \\
\midrule

\parbox[t]{3mm}{\multirow{10}[5]{*}{\rotatebox[origin=c]{90}{Right}}} &
\multirow{3}{*}{Accuracy} &
H &
0.30 & 0.02 & 14.43 & <.001 & \checkmark & -0.01 & 0.02 & -0.23 & .815 & {} \\
{} &
{} &
V &
0.08 & 0.02 & 3.98 & <.001 & \checkmark & -0.01 & 0.02 & -0.24 & .807 & {} \\
{} &
{} &
C &
0.24 & 0.01 & 17.16 & <.001 & \checkmark & -0.01 & 0.02 & -0.31 & .754 & {} \\
\cmidrule{3-13}
{} &
\multirow{3}{*}{Precision} &
H &
-0.00 & 0.00 & -3.48 & .001 & \checkmark & -0.00 & 0.00 & -0.25 & .802 & {} \\
{} &
{} &
V &
-0.00 & 0.00 & -2.45 & .029 & \checkmark & 0.00 & 0.00 & 0.15 & .877 & {} \\
{} &
{} &
C &
-0.00 & 0.00 & -3.96 & <.001 & \checkmark & 0.00 & 0.00 & 0.32 & .747 & {} \\
\cmidrule{3-13}
{} &
\multirow{2}{*}{Lin. slope} &
H &
-0.02 & 0.01 & -3.50 & <.001 & \checkmark & 0.00 & 0.01 & 0.03 & .973 & {} \\
{} &
{} &
V &
-0.02 & 0.01 & -3.37 & .002 & \checkmark & 0.00 & 0.01 & 0.22 & .826 & {} \\
\cmidrule{3-13}
{} &
\multirow{2}{*}{Lin. fit} &
H &
0.03 & 0.11 & 0.29 & 1 & {} & 0.01 & 0.12 & 0.05 & 1 & {} \\
{} &
{} &
V &
0.14 & 0.06 & 2.18 & .059 & {} & 0.06 & 0.07 & 0.85 & .395 & {} \\
\midrule

\parbox[t]{3mm}{\multirow{10}[5]{*}{\rotatebox[origin=c]{90}{Binocular}}} &
\multirow{3}{*}{Accuracy} &
H &
0.01 & 0.02 & 0.61 & 1 & {} & 0.00 & 0.02 & 0.03 & 1 & {} \\
{} &
{} &
V &
0.10 & 0.02 & 4.24 & <.001 & \checkmark & -0.00 & 0.03 & -0.01 & .994 & {} \\
{} &
{} &
C &
0.06 & 0.02 & 3.71 & <.001 & \checkmark & 0.00 & 0.02 & 0.03 & .978 & {} \\
\cmidrule{3-13}
{} &
\multirow{3}{*}{Precision} &
H &
-0.00 & 0.00 & -1.23 & .440 & {} & 0.00 & 0.00 & 0.83 & .440 & {} \\
{} &
{} &
V &
-0.00 & 0.00 & -0.30 & 1 & {} & 0.00 & 0.00 & 0.21 & 1 & {} \\
{} &
{} &
C &
-0.00 & 0.00 & -1.35 & .353 & {} & 0.00 & 0.00 & 0.88 & .380 & {} \\
\cmidrule{3-13}
{} &
\multirow{2}{*}{Lin. slope} &
H &
-0.01 & 0.01 & -1.04 & .598 & {} & 0.00 & 0.01 & 0.43 & .668 & {} \\
{} &
{} &
V &
-0.00 & 0.01 & -0.20 & 1 & {} & 0.00 & 0.01 & 0.05 & 1 & {} \\
\cmidrule{3-13}
{} &
\multirow{2}{*}{Lin. fit} &
H &
0.16 & 0.10 & 1.68 & .187 & {} & 0.05 & 0.11 & 0.43 & .667 & {} \\
{} &
{} &
V &
-0.01 & 0.06 & -0.11 & .915 & {} & 0.07 & 0.06 & 1.15 & .504 & {} \\

\bottomrule
\end{tabular}

    \begin{tablenotes}
    {\footnotesize \item[\ssymbol{1}] z-statistics from the Helmert-style planned comparisons.}
    \end{tablenotes}
    \end{threeparttable}
\end{table*}

% Spatial accuracy
For the left eye, right eye, and binocular signal, as a general matter, the means for spatial accuracy for NO-USC were consistently worse than USC-1 and USC-2 in the horizontal, vertical, and combined directions (see the row of subplots labeled ``Accuracy'' in \autoref{fig:lsmeans}).
These observations were overwhelmingly supported statistically~(\autoref{tbl:stats-calibration}).
The one exception was for the horizontal spatial accuracy of the binocular signal, where there was no significant difference between the uncalibrated and recalibrated data.
The median percent improvement in spatial accuracy from uncalibrated to recalibrated across dimensions and eyes was 12.8\%.
Another consistent finding was that there were not statistically significant differences between calibration methods for accuracy in any case.
This is evident in the row of subplots labeled ``Accuracy'' in \autoref{fig:lsmeans}.

% Spatial precision
For all eyes, spatial precision either did not improve or actually worsened with calibration (see the row of subplots labeled ``Precision'' in \autoref{fig:lsmeans}).
However, spatial precision never significantly worsened with calibration for the binocular signal.
These observations were supported statistically~(\autoref{tbl:stats-calibration}).
Although the effect of calibration on precision was occasionally significant, the median percent change of spatial precision for these statistically significant models was only 0.7\% compared to 12.8\% for accuracy.
So, although some comparisons of spatial precision were statistically significant, the percent change in these cases was trivial.
Another consistent finding was that there were not statistically significant differences between calibration methods for precision in any case.
This is evident in the row of subplots labeled ``Precision'' in \autoref{fig:lsmeans}.

% Linearity slope
Generally, for monocular data (left and right eyes), the linearity slopes were closer to ideal after recalibration for the horizontal and vertical directions (see the row of subplots labeled ``Linearity slope'' in \autoref{fig:lsmeans}).
These observations are supported statistically, except in one case for the vertical linearity slope of the left eye, where there was no significant difference between uncalibrated and recalibrated data~(\autoref{tbl:stats-calibration}).
For binocular data, calibration did not improve linearity slopes.
This is also evident in the row of subplots labeled ``Linearity slope'' in \autoref{fig:lsmeans}.
The median percent change of linearity slope from uncalibrated to recalibrated across dimensions and eyes was 1.5\%.
Another consistent finding was that there were not statistically significant differences between calibration methods for linearity slope in any case.

% Linearity fit
Regarding linearity fit, there was no statistically significant difference between uncalibrated and recalibrated data for any of the eyes (see the row of subplots labeled ``Linearity fit'' in \autoref{fig:lsmeans}).
The median percent change of linearity fit from uncalibrated to recalibrated across dimensions and eyes was -1.3\%.
Another consistent finding was that there were not statistically significant differences between calibration methods for linearity fit in any case.

On balance, calibration improved performance, but there was no evidence that either calibration method was superior.
Summarizing Sect.~\ref{sec:eye-effect} and Sect.~\ref{sec:calibration-effect}, the best overall performance was from the binocular signal calibrated by any means.

\subsection{Analysis of \ac{ET-HMD} versus EyeLink}
\label{sec:results-device}
Our goal here is to compare data quality on the \ac{ET-HMD} and the EyeLink~(see \autoref{tbl:stats-device}).
In the previous section, we found that the binocular signal from the \ac{ET-HMD} always performed at least as well as---and often better than---the left and right eye signals.
Additionally, we found that recalibration improved the vertical and combined spatial accuracy of the binocular signal, but there was no significant difference between the performance of USC-1 and USC-2.
Given these findings, for the purpose of simplification, in this section, we will only present analyses using the binocular signal from the \ac{ET-HMD} for NO-USC and USC-1.
We did, however, assess the effect of Device for each pair of eye and calibration method.
The statistical analysis revealed that the \ac{ET-HMD} had better accuracy and linearity fit than the EyeLink before and after recalibration.

\begin{table*}
    \centering
    %\footnotesize
    %\captionsetup{font=normalsize}
    \caption{
        Comparison of the data quality between the \ac{ET-HMD} (binocular signal) and the EyeLink (left eye signal).}
        % Recalibration was only done for the \ac{ET-HMD}.
        % The F-statistics from the one-way analyses of means are presented.
        % The means for each device are also presented.
        % Note that the means shown have undergone transformations to normality, but these transformations are monotonic and do not affect whether higher or lower values are better.}
    \label{tbl:stats-device}
    \begin{threeparttable}
    \begin{tabular}{@{}cccccccccc@{}}
\toprule
\multirow{2}[2]{2cm}{\centering Calibration method\ssymbol{1}} &
\multicolumn{2}{c}{\multirow{2}[2]{2cm}{\centering Dependent variable}} &
\multicolumn{2}{c}{DF\ssymbol{2}} &
\multirow{2}[2]{*}{F\ssymbol{3}} &
\multirow{2}[2]{*}{p} &
\multirow{2}[2]{*}{Sig.} &
\multicolumn{2}{c}{Means\ssymbol{4}} \\
\cmidrule{4-5}\cmidrule{9-10}

{} & {} & {} & Num & Den & {} & {} & {} & ET-HMD & EyeLink \\
\midrule

\parbox[t]{3mm}{\multirow{10}[5]{*}{\rotatebox[origin=c]{90}{NO-USC}}} &
\multirow{3}{*}{Accuracy} &
H &
1 & 16.95 & 9.00 & .008 & \checkmark & 0.707 & 0.861 \\
{} &
{} &
V &
1 & 17.79 & 4.14 & .057 & {} & 0.755 & 0.889 \\
{} &
{} &
C &
1 & 17.66 & 7.81 & .012 & \checkmark & 0.858 & 1.024 \\
\cmidrule{3-10}
{} &
\multirow{3}{*}{Precision} &
H &
1 & 12.75 & 0.56 & .469 & {} & 0.373 & 0.385 \\
{} &
{} &
V &
1 & 13.21 & 2.57 & .132 & {} & 0.387 & 0.415 \\
{} &
{} &
C &
1 & 11.94 & 1.94 & .189 & {} & 0.473 & 0.509 \\
\cmidrule{3-10}
{} &
\multirow{2}{*}{Lin. slope} &
H &
1 & 19.33 & 1.96 & .177 & {} & 0.990 & 1.009 \\
{} &
{} &
V &
1 & 14.42 & 0.00 & .962 & {} & 1.003 & 1.002 \\
\cmidrule{3-10}
{} &
\multirow{2}{*}{Lin. fit} &
H &
1 & 16.80 & 6.08 & .025 & \checkmark & 6.705 & 5.385 \\
{} &
{} &
V &
1 & 14.76 & 16.89 & <.001 & \checkmark & 5.922 & 3.726 \\
\midrule

\parbox[t]{3mm}{\multirow{10}[5]{*}{\rotatebox[origin=c]{90}{USC-1}}} &
\multirow{3}{*}{Accuracy} &
H &
1 & 19.84 & 7.69 & .012 & \checkmark & 0.697 & 0.861 \\
{} &
{} &
V &
1 & 12.70 & 16.05 & .002 & \checkmark & 0.660 & 0.889 \\
{} &
{} &
C &
1 & 18.22 & 14.36 & .001 & \checkmark & 0.795 & 1.024 \\
\cmidrule{3-10}
{} &
\multirow{3}{*}{Precision} &
H &
1 & 13.16 & 0.33 & .575 & {} & 0.376 & 0.385 \\
{} &
{} &
V &
1 & 12.91 & 2.52 & .137 & {} & 0.387 & 0.415 \\
{} &
{} &
C &
1 & 13.11 & 1.45 & .251 & {} & 0.477 & 0.509 \\
\cmidrule{3-10}
{} &
\multirow{2}{*}{Lin. slope} &
H &
1 & 17.03 & 0.92 & .351 & {} & 0.999 & 1.009 \\
{} &
{} &
V &
1 & 9.83 & 0.02 & .899 & {} & 1.005 & 1.002 \\
\cmidrule{3-10}
{} &
\multirow{2}{*}{Lin. fit} &
H &
1 & 18.73 & 4.25 & .054 & {} & 6.569 & 5.385 \\
{} &
{} &
V &
1 & 14.71 & 17.60 & <.001 & \checkmark & 5.965 & 3.726 \\

\bottomrule
\end{tabular}

    \begin{tablenotes}
    {\footnotesize
        \item[\ssymbol{1}] Recalibration was only done for the \ac{ET-HMD}.
    
        \item[\ssymbol{2}] Estimated numerator and denominator degrees of freedom.
        
        \item[\ssymbol{3}] F-values from the one-way analyses of means.
        
        \item[\ssymbol{4}] The means shown have undergone transformations to normality, but these transformations are monotonic and do not affect whether higher or lower values are better.}
    \end{tablenotes}
    \end{threeparttable}
\end{table*}

\subsection{Temporal precision}
\begin{figure}
    \centering
    \includegraphics[width=\linewidth,trim={0, 0, 0, 0.7cm},clip]{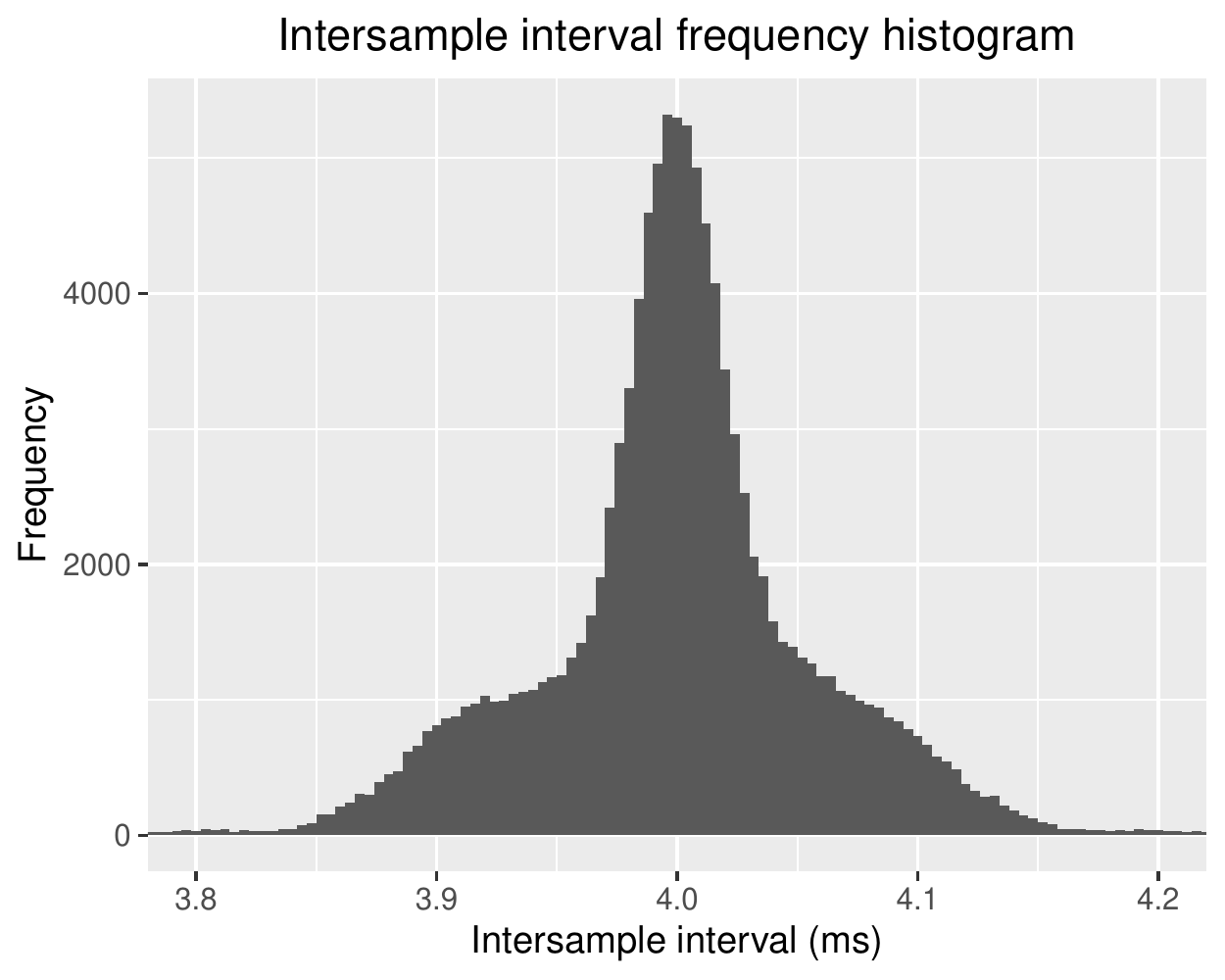}
    \caption{
        A visualization of the temporal precision of the \ac{ET-HMD}, combined across subjects.
	    For the purposes of visualization, \acp{ISI} that are 5\% above or below the expected value are not shown.
	    99.4\% of \acp{ISI} obtained by the \ac{ET-HMD} fit within this range.
	    This means that 99.4\% of the time, the \ac{ET-HMD} has an \ac{ISI} between 3.8 and 4.2~ms.}
    \label{fig:isi-histogram}
\end{figure}

\autoref{fig:isi-histogram} shows the distribution of \acp{ISI} with the \ac{ET-HMD} across subjects.
This distribution has a mean of 4.000~ms and a SD of 0.071~ms.
The \ac{SD} of the distribution represents the temporal precision as we define it.
If the goal is to compute variability irrespective of nominal sampling rate, then one can always compare devices by dividing the SD by the nominal rate.

We also investigated how many samples were dropped with the \ac{ET-HMD}.
We defined a dropped sample as any sample with an ISI over 6~ms (50\% more than the ideal ISI of 4~ms).
In this way, we found that only 4~samples were dropped (1~for one subject, 3~for another) out of 113,264~total samples (approximately 9,400~samples per subject).
We also found that there were 6~extremely short ISIs ($<.04$~ms), each occurring for different subjects.
The cause of these short ISIs is unknown to us.

According to information provided by technical staff at SR-Research, the \ac{ISI} for the EyeLink~1000 is precisely 1~ms.
Also, the Eyelink~1000 does not provide timestamp information with sub-millisecond precision.
Therefore, we are forced to assume that the EyeLink has perfect temporal precision.

%\subsection{Linearity}
% Show the effect of recalibration on linearity with a figure. Plot calibrated slope as thick black line, ideal slope as thinner white dashed line, and precalibrated slope as red line. No samples.

\subsection{Crosstalk}
\begin{figure*}
    \centering
    \includegraphics[width=\linewidth]{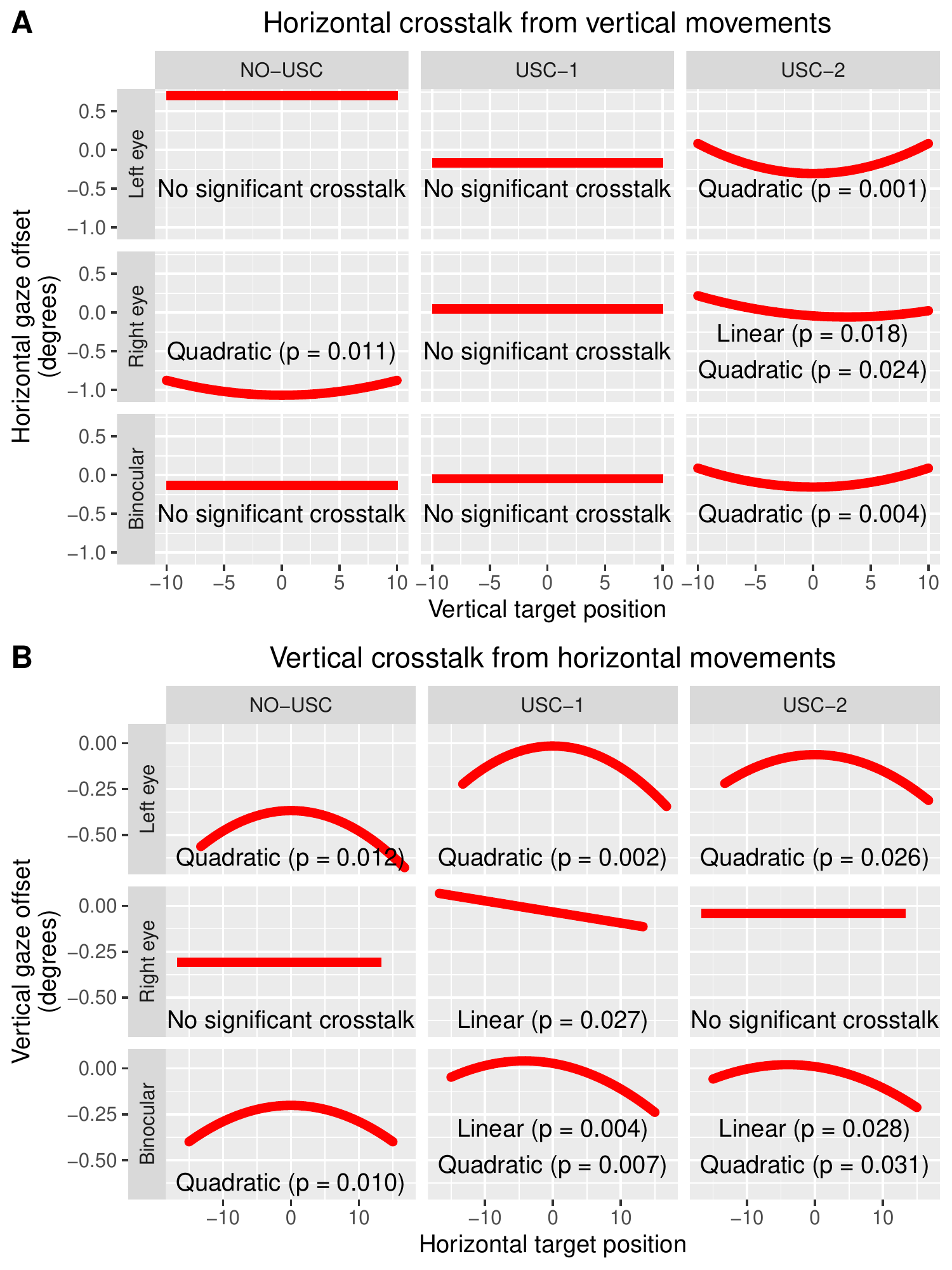}
    \caption{Best-fitting crosstalk models across subjects for each pair of eye and calibration method.
    The p-values shown are the significance of each component according to the stepwise regression.}
    \label{fig:crosstalk-fits}
\end{figure*}

Crosstalk analyses for the \ac{ET-HMD} are presented in \autoref{fig:crosstalk-fits}.
In 7 of 18 test conditions, the best model was the intercept-only model (neither linear nor quadratic components).
In 7 of 18 test conditions, the quadratic-only fit was the best model.
In 3 cases, both a linear component and a quadratic component were needed to optimally fit the data.
In only one case was a linear-only fit optimal.
This is despite the fact that in the literature, only linear fits are tested for and we are not aware of any prior reports of quadratic fits for crosstalk.

For the EyeLink, we found no evidence of significant horizontal or vertical crosstalk.
In other words, in both cases, the model with only the intercept was the best model.

\subsection{Fourier analysis of signals}
\label{sec:fourier-result}
\begin{figure*}
    \centering
    \includegraphics{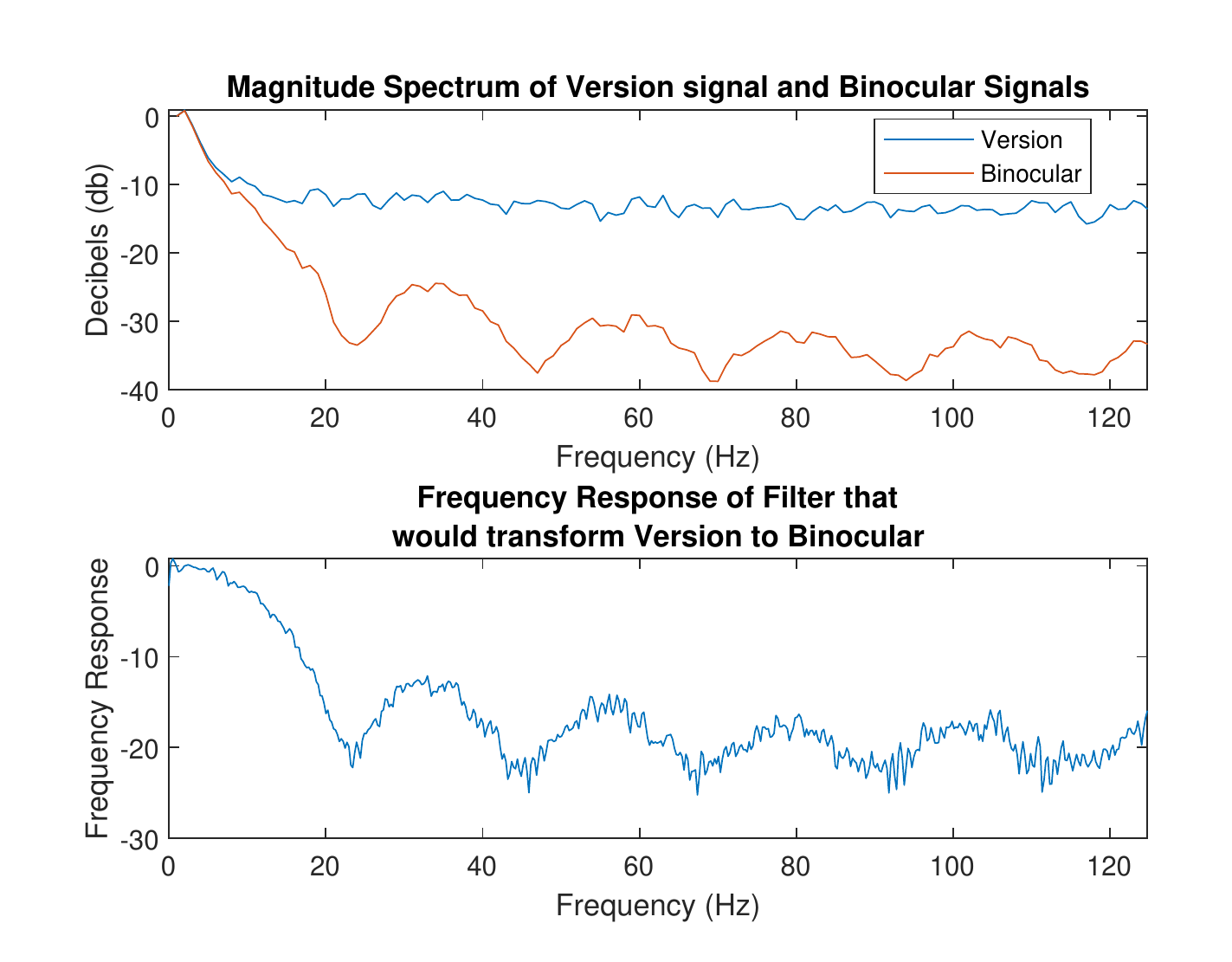}
    \caption{%
        An illustration of the Fourier analysis of signals.
        The top plot is the single-sided magnitude spectra for the version signal and the binocular signal.
        The bottom plot is the frequency response of the filter that would transform the version signal into the binocular signal.
    }
    \label{fig:fourier-results}
\end{figure*}

In the upper panel of Figure \ref{fig:fourier-results}, we can see the magnitude spectra for version signal and the binocular signal.  The binocular signal appears to be a low pass filtered form of the version signal.  The low pass filter is not very sharp (ergo, low-order) and has substantial ringing in the passband.  Using the method suggested by MBaz (see Methods), the lower panel is the frequency response of a filter that would transform our version signal into our binocular signal.  The -3db point is approximately 11 Hz. The -3db point is the point where the power of the signal (magnitude$^2$) is reduced by 50\% or the magnitude is reduced by approximately 30\%.

\begin{figure*}
    \centering
    \includegraphics{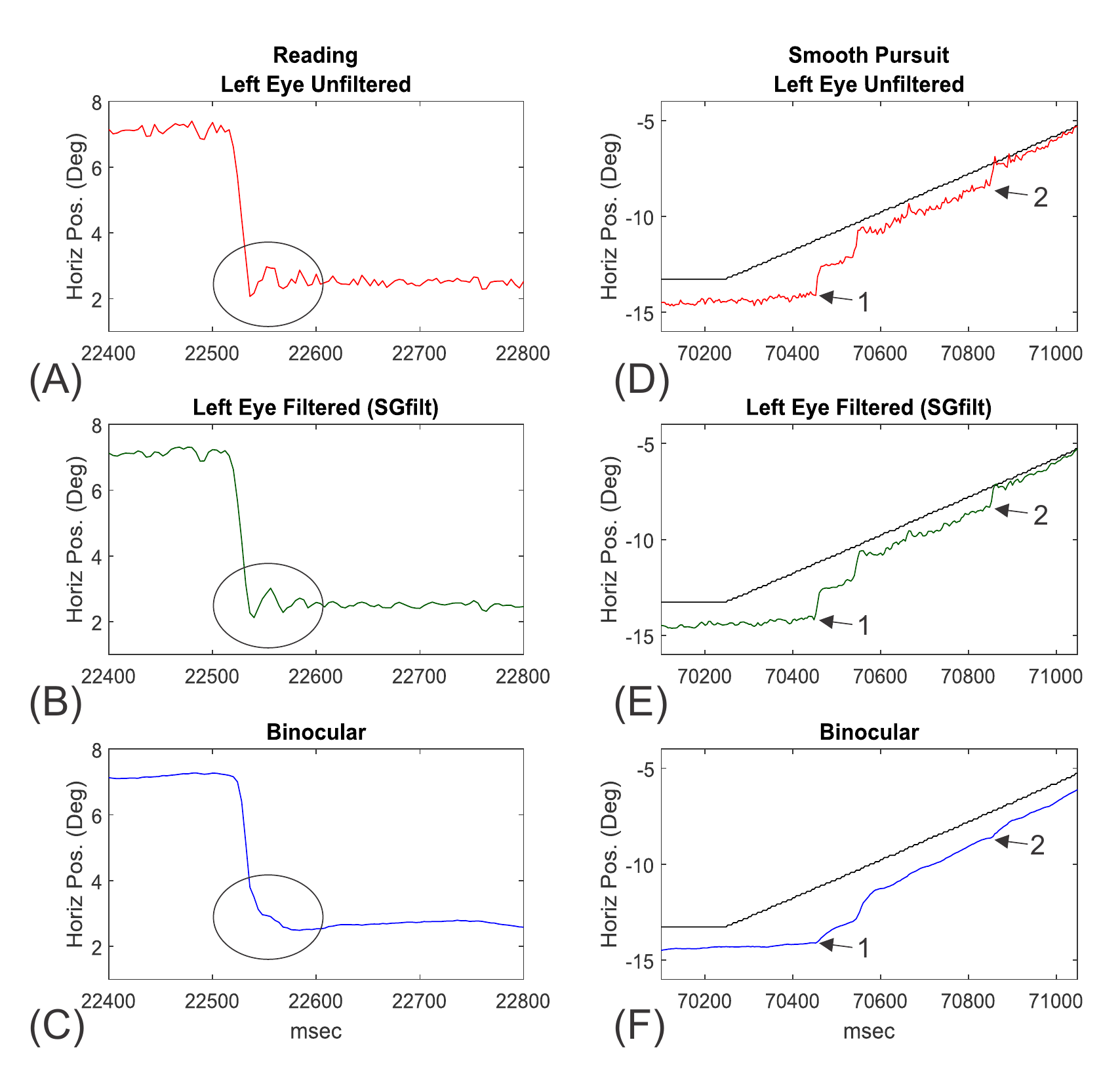}
    \caption{%
        Illustration of saccade features in left eye and binocular signals.
        (A) A saccade of about $4\degree{}$ to the left that is followed by a marked post-saccadic oscillation.
        (B) The same saccade, but filtered with a Savitzky-Golay filter (order 2, window 5).
        (C) The binocular signal.
        Note the complete distortion of the end of the saccade and the eliminiation of the post-saccadic oscillation.
        (D) A section during the initiation of smooth pursuit ramp moving at $10\degree{}$ per second.
        Note the sharp and large initial saccade at the very start of tracking (arrow 1).
        Also note the catch-up saccade around 70850~ms (arrow 2).
        (E) Both saccade features are preserved after applying a Savitzky-Golay filter (order 2, window 5).
        (F) In the binocular signal, both indicated saccades are undetectable as saccades.
        The signals for this figure are from a new data collection.
    }
    \label{fig:saccades-mono-bino}
\end{figure*}

%To get a sense of the effect of such a filter on saccade dynamics, Figure \ref{fig:saccades-mono-bino} (A and B) shows saccades from the version signal on the left (A) and from the binocular signal on the right. These saccades are from the random saccade task.  The smaller saccades in sections C and D, are catch-up saccades from a smooth pursuit task.
To get a sense of the effect of such a filter on saccade dynamics, Figure \ref{fig:saccades-mono-bino} illustrates several saccades unfiltered (A and D), lightly filtered with a Savitzky-Golay filter (B and E), and the binocular signal (C and F).

\clearpage

% The version signal, unfiltered, is too noisy for measuring saccade dynamics. It's better to start with a signal that includes high-frequency components and filter it with a sophisticated filter, rather than starting with a signal (binocular) that has been heavily filtered.

% Top panel: large saccade from random saccades task. Point: the best way to measure saccade dynamics would be to use the filtered version signal as opposed to the binocular signal, because it's obvious that we're losing a lot of information in the binocular signal.

% Bottom panel: show the same thing with smaller saccades (from smooth pursuit).

% Discussion point: what's best depends on your goals. Highly accurate estimate of fixation position -- binocular might be superior. If you want to measure saccade dynamics, binocular signal is a poor choice, and the filtered version signal is a better choice.

\section{Discussion}
\label{sec:discussion}

% Main Findings Summary
We have reported on the eye-tracking quality of SMI's \ac{ET-HMD} in terms of spatial accuracy, spatial precision, temporal precision, linearity, and crosstalk.
We have noted that the binocular signal is a low-pass filtered version of the individual eye signals.
This result affects the choice as to which Vive signal is best for what purpose.
Considering the high accuracy and precision of the binocular signal, it would appear to be the best choice if the goal was simply to identify the position of a fixation (i.e., point of regard (POR) study, or a study emphasizing human-computer interaction).
However, if the goal of the study is to evaluate saccade dynamics, catch-up saccades during smooth pursuit, or the measurement of vergence, the binocular signal simply cannot be used.
For a POR study, across all metrics taken together, the binocular signal with USC-1 recalibration generally would provide optimal performance.
With USC-1, binocular spatial accuracy was $0.382\degree{}$ (H), $0.300\degree{}$ (V), and $0.542\degree{}$ (C).
Binocular spatial precision was $0.054\degree{}$ (H), $0.059\degree{}$ (V), and $0.111\degree{}$ (C).
The temporal precision of the \ac{ET-HMD}, measured as the \ac{SD} of \acp{ISI}, was 0.071 ms.
Binocular linearity slope was 0.999 (H) and 1.005 (V).
Binocular linearity fit was 0.996 (H) and 0.996 (V).
We found statistically significant evidence for crosstalk often having a non-linear form (either quadratic-only or having both linear and quadratic components).

% TODO: Check if these are for artificial eyes -- if so, not fair comparison. Also check if they specify combined or horizontal
The manufacturer-supplied specifications for the \ac{ET-HMD} indicated a typical spatial accuracy of $0.2\degree{}$, but the details of this calculation are not available.
Without recalibration, we measured the combined spatial accuracy of the binocular signal to be $0.671\degree{}$ on average.
Similarly, the manu\-facturer-supplied specifications for the EyeLink~1000 indicated a typical spatial accuracy between $0.25\degree{}$ and $0.50\degree{}$ for human observers.
We measured the combined spatial accuracy of the left eye signal from the EyeLink to be $1.137\degree{}$ on average.
These large differences compared to the manufacturers' claims are part of the reason why data quality analysis is so important.

We made a number of enhancements to a traditional analysis that were effective in providing improved measures of quality.
For our \ac{USC} methods, the recalibration task was prepended to our random saccade task to minimize the time between recalibration and the collection of experimental data.
Recalibration did generally improve spatial accuracy for all three gaze signals (left, right, and binocular) and linearity slopes for the left and right eyes.
Our method for adjusting for saccade latency increased the time that each fixation overlaps with its target, increasing the amount of data available for recalibration and data quality assessment.
Our novel binning procedure appeared to be successful for selecting samples to employ in the recalibration.
Our non-parametric estimate of spatial precision produced measures which were robust to deviations from normality (non-Gaussianity).
Our linearity assessment, including our use of the \ac{R2} to assess the degree of overall linearity and our comparisons of the slope (and confidence limits) to the ideal slope of 1.0, provided new and useful information about the performance of our system.
As far as we are aware, we are the first to assess the fit of crosstalk to both linear and quadratic functions, and we found evidence of a generally non-linear nature of crosstalk inherent in the \ac{ET-HMD}.
% Among the three gaze signal sources available to us, the binocular signal was generally superior.
Our Fourier analysis of the signals was novel and revealed the binocular signal to be a low-pass filtered version of the individual eye signals.
We were the first to precisely characterize the frequency response of such a filter.

% More detail about recalibration
For studies of saccade dynamics, saccades during smooth pursuit, and vergence, monocular signals will need to be employed.
%Based on our overall results, we recommend using the binocular signal over the monocular signals due to improved signal quality.
%However, for the measurement of vergence, accurate left and right eye signals are required.
Our recalibration procedure produced marked enhancements in the accuracy of these monocular signals.
Perfect linearity would mean that the relationship between the measured eye position and the target positions would have a slope of 1.0 and an \ac{R2} of 1.0.
Our recalibration generally brought the slopes of this relationship statistically significantly closer to the ideal slope for the monocular data, but linearity fit was generally unaffected by recalibration.

% Comparison to what others have found/done and EyeLink.
We are not aware of prior reports on the data quality of other eye-tracking \ac{VR} \acp{HMD}, so there is no direct comparison to a similar device we can provide at this time.
We can benchmark the performance of the \ac{ET-HMD} against the much more commonly employed EyeLink~1000.
Spatial accuracy is generally much better with the \ac{ET-HMD} than the EyeLink, especially after recalibration.
Neither spatial precision nor linearity slope are significantly different between the two devices.
Linearity fit is generally significantly better for the \ac{ET-HMD} than the EyeLink.
We found no evidence of significant horizontal or vertical crosstalk in the EyeLink, whereas the \ac{ET-HMD} generally exhibits quadratic crosstalk.

% In Figure XX(a), we show that accuracy is best with the EyeLink, and much poorer with our \ac{ET-HMD} without calibration, but that calibration brings our device much closer to EyeLink's performance.
% Precision is not a problem for our device---even uncalibrated, it has better precision than the EyeLink.
% In Figure XX(b), we show our linearity slope measurements.
% In terms of linearity in the horizontal direction, the best performer is the EyeLink.
% Without recalibration, the \ac{ET-HMD} performs poorly; but with recalibration, our device's performance more closely matches that of the EyeLink.
% In the vertical direction, the EyeLink has worse linearity than in the horizontal direction but is still better than the \ac{ET-HMD} before recalibration.
% With recalibration, the \ac{ET-HMD} surpasses the EyeLink's performance.
% rsquared?

% \begin{figure}
%     \centering
%     \includegraphics{}
%     \caption{Accuracy (x and y), precision (x and y), and linearity (slopes and fit) between the Vive, Vive recalibrated, and EyeLink. Choose the best values from the Vive (between L, R, and B). Four subplots, A-D}
%     \label{fig:my_label}
% \end{figure}

In \autoref{tbl:accuracy-across-devices}, we compare our measures of spatial accuracy with similar measures from 5 different eye trackers as reported by \citet{Blignaut2014}.
Our EyeLink~1000 performs on par with these other machines.
However, the \ac{ET-HMD} performs much better than the other devices without recalibration.
Since our measures of spatial precision, temporal precision, linearity, and crosstalk are unique to the present study, direct comparisons are problematic.
Furthermore, the different recalibration routines applied in the Blignaut study~\citep{Blignaut2014} are different from those employed here, so an accurate direct comparison of calibrated results is not possible.

\begin{table}
    \centering
    %\captionsetup{font=normalsize}
    \caption{Spatial accuracy across various devices, before and after some \ac{USC}.}
    \label{tbl:accuracy-across-devices}
    \begin{threeparttable}
    \begin{tabular}{clcc}
        \toprule
        
        \multirow{2}[2]{*}{Dim} &
        \multirow{2}[2]{*}{Device} &
        \multicolumn{2}{c}{Spatial accuracy\ssymbol{1}} \\
        \cmidrule{3-4}
        
        {} & {} & Before \ac{USC} & After \ac{USC} \\
        
        \midrule
        
        \parbox[t]{3mm}{\multirow{7}{*}{\rotatebox[origin=c]{90}{Horizontal}}} &
        SMI RED 250 (250~Hz)\ssymbol{2} & 0.62 & 0.49 \\
        {} & SMI RED 500 (250~Hz)\ssymbol{2} & 0.69 & 0.44 \\
        {} & SMI RED 500 (500~Hz)\ssymbol{2} & 0.73 & 0.52 \\
        {} & SMI Hi-Speed (500~Hz)\ssymbol{2} & 0.50 & \textbf{0.30} \\
        {} & Tobii TX300 (300~Hz)\ssymbol{2} & 0.79 & 0.31 \\
        {} & EyeLink~1000 (1000~Hz)\ssymbol{3} & 0.68 & - \\
        {} & \ac{ET-HMD} (250~Hz)\ssymbol{3} & \textbf{0.38} & 0.38 \\
        
        \midrule
        
        \parbox[t]{3mm}{\multirow{7}{*}{\rotatebox[origin=c]{90}{Vertical}}} &
        SMI RED 250 (250~Hz)\ssymbol{2} & 0.85 & 0.63 \\
        {} & SMI RED 500 (250~Hz)\ssymbol{2} & 0.62 & 0.47 \\
        {} & SMI RED 500 (500~Hz)\ssymbol{2} & 0.75 & 0.58 \\
        {} & SMI Hi-Speed (500~Hz)\ssymbol{2} & 0.51 & 0.31 \\
        {} & Tobii TX300 (300~Hz)\ssymbol{2} & 0.64 & 0.35 \\
        {} & EyeLink~1000 (1000~Hz)\ssymbol{3} & 0.77 & - \\
        {} & \ac{ET-HMD} (250~Hz)\ssymbol{3} & \textbf{0.47} & \textbf{0.30} \\
        
        \midrule
        
        \parbox[t]{3mm}{\multirow{7}{*}{\rotatebox[origin=c]{90}{Combined}}} &
        SMI RED 250 (250~Hz)\ssymbol{2} & 1.15 & 0.88 \\
        {} & SMI RED 500 (250~Hz)\ssymbol{2} & 1.03 & 0.72 \\
        {} & SMI RED 500 (500~Hz)\ssymbol{2} & 1.17 & 0.87 \\
        {} & SMI Hi-Speed (500~Hz)\ssymbol{2} & 0.80 & \textbf{0.49} \\
        {} & Tobii TX300 (300~Hz)\ssymbol{2} & 1.13 & 0.53 \\
        {} & EyeLink~1000 (1000~Hz)\ssymbol{3} & 1.14 & - \\
        {} & \ac{ET-HMD} (250~Hz)\ssymbol{3} & \textbf{0.67} & 0.54 \\
        
        \bottomrule
    \end{tabular}
    \begin{tablenotes}
        \item[\ssymbol{1}] A bold value indicates the best value in each group.
        
        \item[\ssymbol{2}] The data from these devices come from the study by \citet{Blignaut2014}.
        Their participant-controlled \ac{USC} was used for the ``After USC'' values.
        
        \item[\ssymbol{3}] The data from these devices come from the present study.
        Our USC-1 recalibration method was used for the ``After USC'' values.
    \end{tablenotes}
    \end{threeparttable}
\end{table}

\subsection{How our spatial accuracy method compares to the literature}
We employed the standard definition of spatial accuracy described by \citet{Holmqvist2012}, but our method of choosing which samples to include in our calculation of spatial accuracy differs from that employed in the literature.
We computed spatial accuracy during a random saccade task that spanned $30\degree{}$ horizontally and $20\degree{}$ vertically, where 30 fixations occurred with random durations between 1000-1500 ms.
This choice provided us with 30 fixations, distributed randomly across the field of view, with which to base our accuracy estimate on.
Other researchers typically evaluate accuracy on a smaller subset of calibrations points.
We minimized the impact of saccade latency (see Sect.~\ref{sec:processing}) and used samples in the 400-900~ms interval after each target movement for our calculations.
Gaze samples were screened based on their Euclidean distance to the gaze position centroid in 2 steps.
In the first step, samples with a distance-to-centroid outside Tukey's fences~\citep{Tukey1977} were ignored.
In the second step, samples that were more than $2\degree{}$ away from the centroid were ignored in our calculations.

For comparison, \citet{Tobii2012} used a 9-point grid, presented each target for 2~s, and used data in the 800-1800~ms interval after each target movement.
\citeauthor{Tobii2012} required that 80\% of samples must be valid (non-missing), the \ac{SD} of the sample positions must not exceed $1.5\degree{}$, and the gaze centroid must be no more than $5\degree{}$ away from the target position.
If the set of samples collected for any of the 9~points did not meet these criteria, they recollected the data for those point(s) up to 3~times until the criteria were met.
%% We filter based on distances to centroid. These distances are related to our measure of precision. It would make sense to filter by precision to ensure a relatively stable fixation.
% The basis for the inclusion of a precision filtering threshold is not clear to us.
% It seems to us that such a procedure would confound precision and accuracy.

\citet{Blignaut2014} computed accuracy across 40 targets (8~columns and 5~rows), but used very little outlier screening.
This may have been justified based on their method of determining when fixation occurred.
Fixations were timed based on a mouse click by the subject, which some have concluded results in improved estimation of fixation position~\citep{Nystrom2013}.
\citet{Blignaut2014} required that the gaze centroid must be within $5\degree{}$ from the target position, otherwise the mouse click was not accepted.

\subsection{How our spatial precision method compares to the literature}
\begin{figure}
    \centering
    \includegraphics[width=\linewidth]{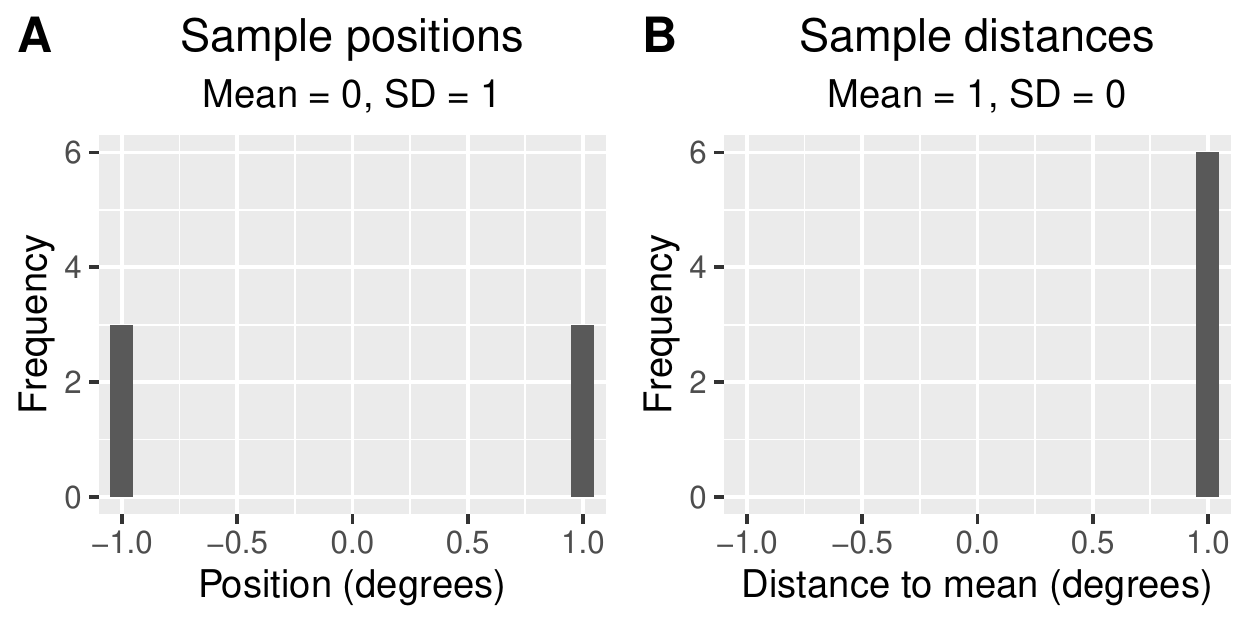}
    \caption{Consider the set of samples with positions $\{-1, -1, -1, 1, 1, 1\}$.
    All samples are a distance of 1~unit away from their central tendency (i.e., 0).
    (A)~The \ac{SD} of sample positions is 1.
    (B)~The \ac{SD} of sample distances is 0.}
    \label{fig:std}
\end{figure}

First, we would like to highlight a certain confusion in the literature on precision.
For example, \citet{Blignaut2012} provide a confusing description of precision calculated using the \ac{SD} approach.
Consider a set of samples that are all a distance of 1~unit away from their central tendency (see \autoref{fig:std} for a simplified example).
In one view, precision is the \ac{SD} of a set of \textbf{\textit{positions}} (see (A) in \autoref{fig:std}).
In the alternative point of view, precision is the \ac{SD} of \textbf{\textit{distances}} (see (B) in \autoref{fig:std}).
\citeauthor{Blignaut2012} are not clear in terms of what they are taking the \ac{SD} of, positions or distances.
They discuss taking ``the standard deviation of a set of points,'' suggesting that they are interested in the \ac{SD} of positions.
But they also provide a \ac{SD} of 0 for points equidistant from their central tendency, which would only be true if they were referring to a \ac{SD} of distances.
The point is, contrary to the claim made by \citet{Blignaut2012}, using Euclidean distance as the distance measure in the \ac{SD} formula does not produce near-zero precision values when the samples are equally spaced around the centroid, unless the \ac{SD} of distances is (wrongfully) used instead of the \ac{SD} of positions.

% In support of this view, \citeauthor{Blignaut2012} state that ``the variance of a set of data samples is a measure of the spread around the mean or central value,'' and they define variance as $\frac{1}{N}\sum_{i=1}^{N}{d_i^2}$, where $d_i$ is some distance measure between an individual sample and the central value.
% They take the square root of variance (the standard deviation) as the measure of precision.
% Following this view, the precision of the set of samples in the above scenario would be 1.0 (see (A) in \autoref{fig:std}).
% In an alternate view, precision is the variance of a set of distances.
% In support of this view, they state that by using Euclidean distance as the distance measure, the standard deviation calculation (which they call SD(E)) would result in a precision near 0 (see (B) in \autoref{fig:std}).
% If they mean to measure the variance of a set of distances, the formula they provided would not be correct.
% The point is, contrary to the claim made by \citet{Blignaut2012}, using Euclidean distance as the distance measure in the standard deviation formula does not produce near-zero precision values when the samples are equally spaced around the centroid, unless the standard deviation of distances is used instead of the standard deviation of positions.

Many, if not most, investigators employ an artificial eye to get machine precision~\citep{Liston2016,Reingold2014,Tobii2012,Wang2017}, and we did not do this.
Many also measure the precision of human eye tracking~\citep{Blignaut2014a,Liston2016,Tobii2012,Wang2017}, as we present here.
\citet{Holmqvist2012} suggest the preferred approach would be to use both artificial eyes and human subjects when assessing spatial precision.
This would allow separate estimates of machine precision and human tracking precision.

The standard measures of spatial precision are \ac{RMS} and the \ac{SD} of positions~\citep{Blignaut2012,Holmqvist2012,Tobii2012}.
\ac{RMS} is often criticized for being heavily dependent on the sampling rate of the eye-tracking device~\citep{Blignaut2012,Tobii2012}.
%\citet{Blignaut2012} point out that the \ac{SD} produces near-zero values when samples are relatively equidistant from their centroid and Euclidean distance is used as the measure of distance.
\citet{Blignaut2012} advocate for the use of a different measure, the \ac{BCEA}, claiming it is easily interpretable, independent of the sampling rate, and independent of the arrangement of samples during a fixation.

We prefer our non-parametric measure of spatial precision, i.e., the median absolute deviation of positions about the median~(MAD).
Like the \ac{BCEA}, the \ac{MAD} is easily interpretable, independent of the sampling rate, and independent of the arrangement of samples during a fixation.
Unlike the \ac{BCEA} which is based on a \ac{SD}, however, the non-parametric \ac{MAD} works well for non-normal distributions.

It should be noted that we could have also used a non-parametric approach to the measurement of spatial accuracy.
However, since nearly all papers that measure spatial accuracy use the same final calculation, we felt that, for the purpose of comparability, we would employ the standard measure.
On the other hand, there are already a number of different methods to calculate spatial precision (\ac{RMS}, \ac{SD}, \ac{BCEA}, and others~\citep{Holmqvist2011}), so we deemed it more appropriate to use what made the most sense to us.

\subsection{How our temporal precision method compares to the literature}
Most investigators use the term ``temporal precision'' to refer to the time between the actual real-time occurrence of an eye movement and the time the movement is registered by the eye-tracking device~\citep{Holmqvist2011}.
In a real-time situation, this would include the exposure time for the digital camera, which is not trivial (on the order of 1~ms), the time to save the image, and the time to process the image to determine the position of the corneal reflection and the pupil.
In the offline case, only the exposure time is relevant.
We do not know the relevant times for the \ac{ET-HMD}.

Herein, we used the term ``temporal precision'' in a different sense.
We used this term to refer to the variation in sample timestamps.
Some eye-tracking devices report a nominal sampling rate and provide timestamps at exactly that sampling rate (e.g., the EyeLink~1000 has a nominal sampling rate of 1000~Hz, and it produces integer timestamps in milliseconds).
% The only variability in this case is due to dropped samples.
However, the \ac{ET-HMD} provides timestamps with nanosecond precision, so we can determine the variability in timestamps to nanosecond precision.
We are aware that the SMI Hi-Speed~1250 tracker also reports timestamps with microsecond precision (see the manually annotated eye-tracker data provided by Marcus Nystr{\"o}m).\footnote{\url{https://www.humlab.lu.se/en/person/MarcusNystrom/}.}
So our sense of temporal precision would apply for that device as well.

\subsection{How our linearity method compares to the literature}
Typically, linearity is assessed by viewing a plot of spatial accuracy as a function of position~\citep{Blignaut2012a,Blignaut2014,Hornof2002}.  
\citet{Reulen1988} employed linear regression to find the best-fitting line where gaze position was the dependent variable and target position was the independent variable.
They found the best-fitting slope and intercept for this relationship, as well as the residuals for each point.
\citeauthor{Reulen1988} then expressed linearity as the maximum residual divided by the range of target positions.
This approach seems to us to be overly influenced by potential outliers or extreme observations.
%Following the notion expressed by Holmqvist et al,
We describe linearity both in terms of the slope of the best-fitting line (and how it compares to the ideal slope of 1.0), and in terms of the degree of fit of the line to the data using the \ac{R2}.

\subsection{How our crosstalk method compares to the literature}
Typically, the assessment of crosstalk assumes linearity of the crosstalk artifact, and crosstalk is expressed as a percentage~\citep{Merchant1967,Reulen1988,Rigas2018}.
To assess crosstalk, we compare 4~models of crosstalk: (1) one with only a linear predictor, (2) one with only a quadratic predictor, (3) one with both linear and quadratic components, and (4) one with the intercept only.
The best-fitting model, found using a stepwise regression, is chosen as the correct model.
In the case of the \ac{ET-HMD}, before our recalibration attempts, none of the best-fitting models were linear-only, and 3 of 6 best-fitting models were quadratic-only (see \autoref{fig:crosstalk-fits}).
We think such an approach is clearly an advance and recommend that others consider it.

\subsection{How our recalibration method compares to the literature}
Some researchers are moving toward participant-controlled calibration routines~\citep{Blignaut2014,Nystrom2013}.
\citet{Nystrom2013} suggest that participant-controlled calibration leads to better estimates of fixation than algorithm-controlled or operator-controlled calibrations.
In the present study, we developed our own sophisticated method for determining which samples are included from each fixation during our algorithm-controlled \acp{USC}.
Future studies comparing different recalibration methods will help determine which approach is best, both in terms of the time needed to perform the \ac{USC} and in terms of how well the spatial data are corrected.
% However, this approach seems to take more time than an algorithm-controlled method.
% So, in the present study, we developed our own sophisticated method for determining which samples are included from each fixation during our \acp{USC}.

\subsection{Limitations}
The \ac{ET-HMD} and its accompanying software are no longer available (neither for purchase, technical support, nor repair), making it difficult for others to replicate our study.

Since we did not have access to an artificial eye, our measures of spatial precision and crosstalk are not measures of the characteristics of the device alone.
Our spatial precision measurements include oculomotor noise, such as drift, tremor, and microsaccades.
Our crosstalk measurements are sensitive to crosstalk at both the device and biological levels.

If we had studied the same subjects on the \ac{ET-HMD} and the EyeLink~1000, we would have had a much more powerful and meaningful test of the comparison between these devices.
Although, for our analysis, a random saccade task was used on both devices, spanning $30\degree{}$ horizontally and $20\degree{}$ vertically, there were other technical differences which could be minimized in future studies.
For example: (1) the targets, (2) the background color and target color, (3) viewing distance, (4) presence of a \acl{USC} prior to the main task, and (5) whether subjects were financially and/or academically motivated to participate (yes for the EyeLink, no for the \ac{ET-HMD}).

% As other studies have done, we would like to assess the linearity of spatial accuracy and spatial precision, rather than just the linearity of gaze position.

\subsection{Future directions}
In the future, we want to look into various options for an artificial eye, including a moving artificial eye.
Mounting such an artificial eye within the \ac{ET-HMD} would present a challenge.
We would also like to record the same people on the \ac{ET-HMD}, the EyeLink~1000, and even other devices to provide a more comprehensive evaluation across more devices.
For quality control among eye-tracking \ac{VR} \acp{HMD}, new measures of the quality of vergence signals and of the impact of the \ac{VAC} should be developed.
As numerous studies have shown, the \ac{VAC} negatively impacts oculomotor function (phoria)~\citep{Paulus2017,Shibata2011}, causes visual discomfort~\citep{Shibata2011}, leads to ``simulator sickness'' including headaches and nausea~\citep{Hakkinen2006,Huang2015}, increases visual fatigue~\citep{Hoffman2008,Shibata2011}, and reduces cognitive performance~\citep{Daniel2019}.
Therefore, it is reasonable to believe that the \ac{VAC} would also negatively impact data quality.
Various efforts are being made to reduce or outright eliminate the \ac{VAC} in \ac{VR} devices~\citep{Huang2015,Kramida2016}.
But until these solutions become widely present in eye-tracking \ac{VR} \acp{HMD}, quantifying the impact of the \ac{VAC} could be useful for future studies.

\section*{Acknowledgements}
This material is based upon work supported by the National Science Foundation Graduate Research Fellowship under Grant No.~DGE-1144466.
The study was also funded by 3 grants to Dr. Komogortsev: (1) National Science Foundation, CNS-1250718 and CNS-1714623, \url{www.NSF.gov}; (2) National Institute of Standards and Technology, 60NANB15D325, \url{www.NIST.gov}; (3) National Institute of Standards and Technology, 60NANB16D293.
Any opinions, findings, and conclusions or recommendations expressed in this material are those of the author(s) and do not necessarily reflect the views of the National Science Foundation or the National Institute of Standards and Technology.


% Change the meaning of \VON to use the prefix for the bibliography
\DeclareRobustCommand{\VON}[3]{#3}

%\bibliographystyle{plainnat}
%\bibliographystyle{apacite}
%\bibliography{data-quality-nourl}   % name your BibTeX data base

%\begin{thebibliography}{}
%%
%% and use \bibitem to create references. Consult the Instructions
%% for authors for reference list style.
%\end{thebibliography}

\end{document}